\documentclass[10pt,twocolumn,twoside]{IEEEtran}
\usepackage{amssymb, amsmath}
\usepackage{bm}
\usepackage{color}
\usepackage{graphicx}
\usepackage{cite}
\usepackage{booktabs}
\usepackage{multirow}
\usepackage{threeparttable}
\usepackage{enumerate}
\usepackage[english]{babel}
\usepackage{algorithmicx}
\usepackage{algorithm}
\usepackage{algpseudocode}
\usepackage{nomencl}
\usepackage{commath}
\usepackage{makecell}
\usepackage{blindtext}
\usepackage{soul}
\usepackage{comment}
\usepackage{mathtools}
\usepackage[svgnames]{xcolor}
\usepackage[caption=false,font=footnotesize]{subfig}
\usepackage{tikz}
\usepackage{derivative}
\usepackage[export]{adjustbox}
\usepackage[siunitx, RPvoltages]{circuitikz}
\usetikzlibrary{positioning, shapes}
\usepackage{url}
\usepackage{accents}
\usepackage[inter-unit-product =\cdot]{siunitx}

\newcommand{\vect}[1]{\mathbf{#1}}
\newcommand{\matr}[1]{\mathbf{#1}}
\newcommand{\tran}{\mathsf{T}}
\newcommand{\herm}{\mathsf{H}}

\newcommand{\Yan}[1]{{\color{magenta}[Yan: {#1}]}}

\DeclareMathOperator*{\dist}{dist}

\DeclareMathOperator*{\col}{col}
\DeclareMathOperator*{\diag}{diag}
\newcommand\eqm[1]{\bar{#1}}

\DeclareFontFamily{U}{mathx}{\hyphenchar\font45}
\DeclareFontShape{U}{mathx}{m}{n}{
      <5> <6> <7> <8> <9> <10>
      <10.95> <12> <14.4> <17.28> <20.74> <24.88>
      mathx10
      }{}
\DeclareSymbolFont{mathx}{U}{mathx}{m}{n}
\DeclareFontSubstitution{U}{mathx}{m}{n}
\DeclareMathAccent{\widecheck}{0}{mathx}{"71}
\DeclareMathAccent{\wideparen}{0}{mathx}{"75}

\newtheorem{lemma}{Lemma}
\newtheorem{definition}{Definition}
\newtheorem{remark}{Remark}
\newtheorem{proposition}{Proposition}

\makeatletter
\renewcommand*\env@matrix[1][\arraystretch]{%
  \edef\arraystretch{#1}%
  \hskip -\arraycolsep
  \let\@ifnextchar\new@ifnextchar
  \array{*\c@MaxMatrixCols c}}
\makeatother

\title{
Power System Electromagnetic Transient Stability: an Analysis Based on Convergent Hamiltonian
}
\author{Xinyuan Jiang, Constantino M. Lagoa, and Yan Li% <-this % stops a space
\thanks{The authors are with the Department of Electrical Engineering, The Pennsylvania State University, University Park, PA 16802, USA (e-mail: xuj49@psu.edu; cml18@psu.edu; yql5925@psu.edu).}}

\begin{document}

\maketitle

\begin{abstract}
Transient stability is crucial to the reliable operation of power systems. Existing theories rely on the simplified electromechanical models, substituting the detailed electromagnetic dynamics of inductor and capacitor with their impedance representations. However, this simplification is inadequate for the growing penetration of fast-switching power electronic devices. Attempts to extend the existing theories to include electromagnetic dynamics lead to overly conservative stability conditions.
To tackle this problem more directly, we study the condition under which the power source and dissipation in the electromagnetic dynamics tend to balance each other asymptotically. This is equivalent to the convergence of the Hamiltonian (total stored energy) and can be shown to imply transient stability. Using contraction analysis, we prove that this property holds for a large class of time-varying port-Hamiltonian systems with (i) constant damping matrix and (ii) strictly convex Hamiltonian. Then through port-Hamiltonian modeling of the electromagnetic dynamics, we obtain that the synchronized steady state of the power system is globally stable if it exists. This result provides new insights into the reliable operation of power systems. 
The proposed theory is illustrated in the simulation results of a two-machine system.
%The insight from this result is that the major challenge in the reliable operation of future power systems is the existence of a synchronized steady state, more so than its stability. %The theory is illustrated in detailed analysis of the simulated numerical example.
\end{abstract}

\section{Introduction}

\PARstart{T}{ransient} stability is crucial to the planning and operation of power systems and has become even more so with the increasing penetration of power electronic technologies, whose dynamics are dependent primarily on their local controls. These local controls operate on the electromagnetic time scale from 1 \unit{\micro s} to 1 \unit{\milli s}, which necessitates more detailed analyses into the transient dynamics of power systems in the electromagnetic time scale~\cite{hatziargyriou2020definition}. In addition to the increased complexity in the types of dynamic agents involved, there is also a growing trend of individual generators for industrial and commercial power systems to sustain islanding situations during blackouts and to sell power to the bulk power system during normal operation~\cite{dai2024practical}. This creates increased complexity in the loading and configuration, which magnifies the limitations of the existing methods for transient stability analysis.
There are mainly four types of methods for analyzing transient stability, namely, angle stability, feedback linearization, passivity, and direct methods. 

Firstly, angle stability~\cite{bera2021identification,dai2024practical} aims at monitoring the torque angles of every synchronous generator in the system to detect whether some generators are entering into stable or unstable swing post fault. It relies on the observation that if the angles do not exceed the limits of the power-angle curves, in out-of-step conditions~\cite{tziouvaras2004out}, then the restoring and damping torque cause the angles to synchronize. The limitation, however, is that it does not study the system dynamics directly, but aims to extract patterns from numerical simulations of instability. Therefore, it does not provide enough insight into the effects of the controls.

Secondly, feedback linearization can algebraically linearize and decouple the interactions between the generators and the loads to obtain linear subsystems which  are suitable for linear control. In~\cite{mahmud2013partial}, an synchronous generator excitation control is designed to guarantee transient stability of the simplified multimachine power system model. In~\cite{bidram2013distributed}, a microgrid secondary control is designed for improving the tracking performance for frequency and voltage. However, the limitation is that the electromagnetic dynamics usually cannot be fully linearized so that only partial stability can be attained.

Thirdly, passivity is a distributed subsystem-level property for verifying overall transient stability. It is closely related to physical energy dissipation, and so it can handle systems with nonzero line conductance~\cite{spanias2018system,siahaan2024decentralized}. The main issue in its application to transient stability is that the nonlinear transformations between different $dq$-reference frames of the generators is a major obstacle for verifying passivity~\cite{caliskan2014compositional}.

Last but not least, the direct method for transient stability analysis~\cite{varaiya1985direct,chiang1987foundations,rimorov2018approach,cheng2021transient} consists of (i) a model of the power system with simplifications, and (ii) the associated energy function (a function of voltage angles and frequencies), which decreases in time monotonically. Under certain conditions, the energy function becomes locally a Lyapunov function, but their differences are seen from the following aspects:
\begin{itemize}
    \item To define the energy function, the periodic angle domain is unrolled into a real domain.\footnote{We mean that the $2\pi$ equivalence of the domain $\mathbb{T}^n$ is removed so that two equivalent angle vectors in $\mathbb{T}^n$ as their difference is multiplies of $2\pi$ are no longer deemed equivalent~\cite{schiffer2019global,forni2014differential}.}  
    This explains why the input power, which is non-conservative, can seemingly be ``integrated'' to get a linear function of the unrolled angles, i.e.,
    \begin{equation*}
        \int P_m \Delta \omega\, dt = P_m \delta.
    \end{equation*}
    Then, by designating a reference bus of $0$ voltage angle, the steady state becomes an isolated equilibrium point.\footnote{If a reference bus is not chosen, the system has angle symmetry, which causes a set of equilibrium points whose angles are rigidly shifted. To deal with this problem, usually a reference angle rotating at the average frequency is chosen~\cite{varaiya1985direct}. However, this does not work without the constant impedance assumption.} %Given that this equilibrium point is locally asymptotically stable,\footnote{Without this condition, it is almost impossible to establish that the region of attraction is nonempty. See the same idea used later in~\cite{schiffer2014conditions}.} the goal is to find the region of attraction of this equilibrium point.
    \item The stability of the equilibrium point is not implied from the energy function because of its boundedness. Instead, given that this equilibrium point is locally asymptotically stable,\footnote{Without this condition, it is almost impossible to establish that the region of attraction is nonempty. See the same idea used later in~\cite{schiffer2014conditions}.}
    an estimated region of attraction can be obtained by calculating the values of the energy function at those unstable equilibrium points located on the boundary of the region of attraction. Then the minimum of the function values is used for estimating the critical clearly time for a fault given the post-fault initial condition~\cite{cheng2021transient}.
\end{itemize}

The limitation of the direct method, which prevents its wider application, is the difficulty in finding the unstable equilibrium points on the boundary of the region of attraction. It is worth noting that the estimated region of attraction from the direct method is unbounded~\cite{chiang1987foundations} compared to the bounded estimate from angle stability~\cite{dai2024practical}.

%The task to find all the unstable equilibrium points on the stability boundary is the main obstacle for applying the energy function method in practice. Moreover, the theory assumes that limit cycles do not exist on the stability boundary, which is true for the simplified model but prohibitively difficult to verify for a more detailed model.

An important limitation that is shared by almost all existing methods is that they inadvertently focus on the slower electromechanical dynamics. In an electromechanical model, the dynamics of the inductors and capacitors are replaced by the impedance value at a certain frequency. In contrast, in the full-order equations for the inductor and capacitor in (\ref{E:inductor}) and (\ref{E:capacitor}) in a stationary reference frame, there is a lack of frequency-dependent terms. This is because the steady-state frequency, in fact, an incidence of an energy balance in the system. For example, the value of the steady-state frequency changes as soon as the load is changed slightly. Therefore, based on the electromagnetic model, the equilibrium of the system is not an equilibrium point, but rather an energy-balancing limit cycle. If an infinite bus is present, the frequency of the limit cycle is equal to the frequency of the infinite bus. If an infinite bus is not present, the frequency of the limit cycle is determined autonomously.

By viewing the power system transient stability problem as checking the ability of the system to reach a certain energy balance, it is reasonable to ask the question: \emph{Whether there is a certain class of systems such that its Hamiltonian (total stored energy) is convergent to the same value (not necessarily zero) along every solution?}

%the existence of a limit cycle of a certain constant Hamiltonian (total stored energy) implies that the Hamiltonian along every solution converges to the Hamiltonian at the limit cycle.} %In other words, whether the Hamiltonian would tend to a given balance at the limit cycle as the internal energy source and dissipation  eventually cancel each other.

It will be shown in this paper that the answer to the above question is affirmative. In particular, the class of time-varying port-Hamiltonian system with constant damping matrix and strictly convex Hamiltonian verifies that the Hamiltonian along every trajectory converges to the same value (the converging Hamiltonian principle in Proposition~\ref{prop_convergence}). %Note that this is not obvious because it is possible for a dissipative time-varying system energy source to enter into an oscillatory Hamiltonian. 
It is implied from a more fundamental property called contraction in the quotient space, which is developed first in Section~\ref{sec_contraction}. 
Then, by modeling the electromagnetic dynamics of the power system as a time-varying port-Hamiltonian system in Section~\ref{sec_stability}, we obtain converge of the Hamiltonian, which combined with the constant (not depending on the state) network structure of the power system yields the global attractivity of the limit cycle. %This result is consistent with a previous result on the global stability of the electromechanical model~\cite{schiffer2019global}. 
Finally, in Section~\ref{sec_test}, electromagnetic simulation of a two-machine power system from random initial conditions confirms the theoretical results, and the existence of a synchronized limit cycle is identified as the main challenge in operating AC power systems.
The contributions of this paper are summarized as follows:

\begin{itemize}
    \item The global attractivity of the limit cycle steady state of the electromagnetic power system model is proved based on the converging Hamiltonian principle. The compositional feature of the method provides a general framework for cooperation of synchronous generator and power electronics control.
    \item A large class of time-varying port-Hamiltonian system is proved to be contractive in a special quotient space. Only the constant positive-definiteness of the damping matrix and the strict convexity of the Hamiltonian are assumed, which enables its wider application.
    \item The numerical example shows that several instability concepts in power engineering are related to the nonexistence of a synchronized limit cycle, rather than related to the non-attraction of its orbit.
\end{itemize}

%The rest of this paper is structured as follows. 
%Section~\ref{sec_background} contains backgrounds on matrix measure and pH system. Section~\ref{sec_contraction} contains the definition and theoretical results on contraction in the quotient space. Section~\ref{sec_stability} contains a detailed analysis of the electromagnetic model of a two-machine power system. Section~\ref{sec_test} provides a simulated case study. Section~\ref{sec_conclusion} is the conclusion.
%\input{preliminaries}
%\input{theory}

{\it Notation:}
The imaginary unit is $j$. %The matrix form of $j$ is defined as $\matr J = [0, -1;1, 0]$.
A vector of zeros is $0_n$. A vector of ones is $1_n$. Their subscripts are usually omitted when the dimension is clear. 
%The real and imaginary part of $A \in \mathbb{C}$ are $\Re\{A\}$ and $\Im\{A\}$.
%An identity matrix is $\matr I_n$. A zero matrix is $\matr 0_{n\times m}$. The subscript of these notations is sometimes omitted if the dimension is clear from the context.
%The $1$-torus with $2\pi$ equivalence is $\mathbb{T} = \mathbb{R}/2\pi\mathbb{Z}$~\cite{FB-ADL:04}. 
The transpose of a matrix is $\matr A^\tran$; the Hermitian transpose is $\matr A^\herm$. For a matrix $\matr A$ in an inner product space, the adjoint is $\matr A^*$. %; $\matr A$ is self-adjoint if $\matr A = \matr A^*$; and $\matr A$ is skew-adjoint if $\matr A = -\matr A^*$. %; and the pseudoinverse is $\matr A^\dagger$. 
The symbol $\col(\vect x_i)$ denotes a column vector that stacks the vectors $\vect x_i$ for $i = 1,\ldots,\, n$. 
%The inner product we throughout this paper assume is $\langle\vect y, \vect x\rangle = \Re\{\vect y^* \vect x\}$ for $\vect x, \vect y \in \mathbb{C}^n$. %, where $\mathbb{C}^n$ is a vector space over the field $\mathbb{R}$.
%This inner product can be characterized as
%\begin{equation} \label{E:complex_inner}
%    \langle \vect y, \vect x \rangle = \langle \Re\{\vect y\}, \Re\{\vect x\}\rangle + \langle \Im\{\vect y\}, \Im\{\vect x\}\rangle
%\end{equation}
%where the inner products in the RHS are real inner products.
%The adjoint of a linear mapping $\matr L: \mathbb{C}^n \to \mathbb{C}^n$ is $\matr L^*$.
%The Hermitian part of a matrix $\matr A$ is $\mathrm{He}\{\matr A\} = \frac{1}{2} (\matr A + \matr A^*)$.
%For a Hermitian matrix $\matr L$, $\matr L > 0$ denotes positive definiteness, i.e., $\langle \vect x, \matr L \vect x\rangle < 0$ for all $\vect x \in \mathbb{C}^n$.  
The complex vector space $\mathbb{C}^n$ we consider in this paper is equivalent to the $\mathbb{R}^{2n}$ considering the mapping 
\begin{equation*}
    \mathcal U: \vect x \mapsto \hat{\vect x} = \col(\big[\Re x_i,\, \Im x_i \big]^\tran). 
\end{equation*}
The derivative of real-valued functions defined on $\mathbb{C}^n$ are taken after mapping them to the equivalent real function. For $\vect f: \mathbb{C}^n \to \mathbb{C}^n$, the real Jacobian $D \vect f(\vect x)$ is the Jacobian of $\vect f(\mathcal U^{-1}(\hat{\vect x}))$. For $H: \mathbb{C}^n \to \mathbb{R}$, the real Hessian $D^2 H(\vect x)$ is the Hessian of $H(\mathcal U^{-1}(\hat{\vect x}))$. 
An exception is the complex gradient. For the inner product $\langle \vect z, \vect x \rangle = \Re\{\vect z^\herm \vect x\}$, the complex gradient is defined, for $H: \mathbb{C}^n \to \mathbb{R}$, as $\nabla H(\vect x) = 2 \col(\frac{\partial H}{\partial x_i^*})$, where $\frac{\partial H}{\partial x_i^*}$ is the Wirtinger derivative~\cite{remmert1991theory}. The complex gradient allows us to replace a direction derivative by an inner product: 
\begin{equation*}
    \frac{\partial}{\partial \vect x} H(\vect x)\cdot \vect v = \Re\{\nabla H(\vect x)^\herm \vect v\}.
\end{equation*}

\section{Background} \label{sec_background}

In this section, we discuss the background for the main theoretical development in Section~\ref{sec_contraction}: the contraction property implied from the port-Hamiltonian structure of a nonlinear time-varying (NLTV) system. We introduce this background knowledge by reviewing standard results on contraction of a general NLTV system, and contraction of a port-Hamiltonian NLTV system. We highlight the conservativeness of the standard results to motivate the more involved theoretical development in Section~\ref{sec_contraction}.

\subsection{Standard Contraction Definition}
Consider the NLTV system $\dot{\vect x} = \vect f(t, \vect x)$ for $t \in \mathbb{R}$ and $\vect x \in \mathbb{D}$, for an invariant set $\mathbb{D} \subset \mathbb{C}^n$ such that, at each $\vect x \in \mathbb{D}$, the tangent space\footnote{The tangent space $T_{\vect x} \mathbb{D}$ of a manifold $\mathbb{D}$ at $\vect x$ is the collection of tangent vectors of smooth curves in $\mathbb{D}$ that passes through $\vect x$; see~\cite[Def.~3.33]{FB-ADL:04}.} is $T_{\vect x} \mathbb{D} = \mathbb{C}^n$ with the inner product 
\begin{equation} \label{E:inner_product}
    \langle \vect y, \vect x \rangle = \Re\{\vect y^\herm \matr P \vect x\},
\end{equation}
where $\matr P \in \mathbb{C}^{n\times n}$ is Hermitian positive-definite. This inner product verifies the inner product axioms for the vector space $\mathbb{C}^n$ with the field $\mathbb{R}$. The associated norm is
\begin{equation*}
    \|\vect x\| = \sqrt{\langle \vect x, \vect x \rangle} = \sqrt{\Re\{\vect x^\herm \matr P \vect x\}}.
\end{equation*}
%This space is equivalent to $\mathbb{R}^{2n}$.
%By this definition, the inner product space $\mathbb{C}^n$ considered is equivalent to $\mathbb{R}^{2n}$ by observing that
%\begin{equation*}
%    \langle \vect y, \vect x \rangle_{\mathbb{C}^n} = \Re\{\vect y^\herm \vect x\} = \Re\{\vect y\}^\tran \Re\{\vect x\} + \Im\{\vect y\}^\tran \Im\{\vect x\}.
%\end{equation*}
We choose to work with the space $\mathbb{C}^n$ instead of the equivalent real space $\mathbb{R}^{2n}$ because complex variables readily represent amplitudes and angles in models of the power system. 

%Note that the default inner product definition is not the only valid one verifying the inner product axioms. Given a Hermitian positive-definite matrix $\matr P = \matr P^\herm$, we can verify that $\langle \vect y, \vect x \rangle = \Re\{\vect y^\herm \matr P \vect x\}$ is another valid inner product. 

Contraction of the NLTV system is a sufficient condition for stability that requires two solutions from any two initial conditions to converge exponentially to each other in terms of the norm of their difference in time. The contraction property is particularly useful in stability analysis in that the global condition can be checked locally with the Jacobian matrix $D \vect f(t, \vect x)$: the NLTV system is said to be infinitesimally contracting if, for some $c > 0$ (the contraction rate),
\begin{equation} \label{E:infinitesimal}
     \langle \delta, D \vect f(t, \vect x) \delta \rangle \leq -c \langle \delta, \delta \rangle,
\end{equation}
for all $t \in \mathbb{R}$ and $\delta \in T_{\vect x} \mathbb{D} = \mathbb{C}^n$.
It is not hard to prove that if the domain $\mathbb{D}$ is convex, then infinitesimal contraction implies global contraction~\cite{sontag2010contractive,simpson2014contraction}; that is, for every $\vect x_1, \vect x_2 \in \mathbb{D}$ and $t_0 \in \mathbb{R}$, the condition (\ref{E:infinitesimal}) implies
\begin{equation} \label{E:distance_contraction}
    \big\|\Phi(t, t_0, \vect x_1) - \Phi(t, t_0, \vect x_2)\big\| \leq e^{-c (t - t_0)} \big\|\vect x_1 - \vect x_2\big\|
\end{equation}
where $\Phi(t, t_0, \vect x_0)$ is the solution of the NLTV system from the initial condition $(t_0, \vect x_0)$. The contraction condition (\ref{E:infinitesimal}) depends critically on the inner product chosen. Let us examine the condition (\ref{E:infinitesimal}) in more detail.

For every matrix $\matr A \in \mathbb{C}^{n\times n}$, the matrix measure is defined as\footnote{The matrix measure is commonly defined in terms of a vector norm~\cite{soderlind2006logarithmic} or in terms of an inner product. The latter is more appropriate for port-Hamiltonian systems, which are naturally defined in inner product spaces.}
\begin{equation} \label{E:matrix_measure}
    \mu(\matr A) = \sup_{\delta \in \mathbb{C}^n \backslash \{0\}} \frac{\langle \delta, \matr A \delta \rangle}{\langle \delta, \delta \rangle}.
\end{equation}
From this definition, the contraction condition (\ref{E:infinitesimal}) is equivalently expressed as $\mu(D \vect f(t, \vect x)) \leq -c$. 
Since the skew-adjoint part of $\matr A$ yields zero in the numerator of (\ref{E:matrix_measure}), $\mu(\matr A)$ is equal to
\begin{equation} \label{E:self-adjoint}
    \mu(\matr A) = \sup_{\delta \in \mathbb{C}^n \backslash \{0\}} \frac{\langle \delta, \frac{1}{2}(\matr A + \matr A^*) \delta \rangle}{\langle \delta, \delta \rangle}
\end{equation}
%The definition of adjoint differs depending on the definition of inner product chosen.
%If the inner product is $\langle \vect y, \vect x \rangle = \Re\{ \vect y^\herm \vect x \}$, then the adjoint of $\matr A$ is $\matr A^\herm$. If the inner product is $\langle \vect y, \vect x \rangle = \Re\{ \vect y^* \matr P \vect x \}$, then 
The adjoint $\matr A^*$ is a mapping such that $\langle \vect x, \matr A \vect y \rangle = \langle \matr A^* \vect x, \vect y \rangle$; see~\cite[Sec.~4.4]{bressan2012lecture}.
For the inner product (\ref{E:inner_product}), the adjoint of $\matr A$ is obtained as $\matr P^{-1} \matr A^\herm \matr P$, which we substitute into (\ref{E:self-adjoint}) to get
\begin{align*}
    \mu(\matr A) &= \sup_{\delta \in \mathbb{C}^n \backslash \{0\}} \frac{\Re\big\{ \delta^\herm \matr P \frac{1}{2} (\matr A + \matr P^{-1} \matr A^\herm \matr P) \delta \big\}}{\Re\{ \delta^\herm \matr P \delta \}} \\
    &= \sup_{\delta \in \mathbb{C}^n \backslash \{0\}} \frac{\Re\big\{\frac{1}{2} \delta^\herm (\matr P \matr A + \matr A^\herm \matr P) \delta \big\}}{\Re\{ \delta^\herm \matr P \delta \}} \\
    &= \sup_{s \in \mathbb{C}^n \backslash \{0\}} \frac{\Re\big\{\frac{1}{2} s^\herm \matr P^{-\frac{1}{2}} (\matr P \matr A + \matr A^\herm \matr P)\matr P^{-\frac{1}{2}} s \big\}}{\Re\{ s^\herm s \}},
\end{align*}
where $s = \matr P^{\frac{1}{2}} \delta$.
We then obtain the following expression for the matrix measure that is dependent on the matrix $\matr P$:
\begin{equation*}
    \mu(\matr A) = \lambda_{\max} \bigg\{ \frac{1}{2} \matr P^{-\frac{1}{2}} (\matr P \matr A + \matr A^\herm \matr P) \matr P^{-\frac{1}{2}} \bigg\}.
\end{equation*}
By~\cite[Thm.~4.6]{Khalil:1173048}, if $\matr A$ is Hurwitz, there exist (many) $\matr P$'s and the associated inner products such that $\mu(\matr A) < 0$. However, not all $\matr P$'s and such inner products verify $\mu(\matr A) < 0$. 

In dealing with the general NLTV systems, the skew-adjoint part of the Jacobian matrix is ignored in order to check contraction. The skew-adjoint part can usually be related to the energy-preserving or structural part of the dynamics; this part is explicitly separated from the energy-dissipating or damping part in the port-Hamiltonian formulation of NLTV systems. %As will be shown, this explicit separation leads to a more befitting contraction property enjoyed by the port-Hamiltonian system in Proposition~\ref{prop_main}.

\subsection{Port-Hamiltonian System} \label{sec_pH}
In this paper, we consider the input-state-output port-Hamiltonian (pH) system with a constant damping matrix:
\begin{equation} \label{E:pH} 
    \Sigma: \begin{cases}
        \dot{\vect x} = (\matr J(t, \vect x) - \matr R) \nabla_{\vect x} H(t, \vect x) + \matr G \vect u \\
    \vect y = \matr G^\herm \nabla_{\vect x} H(t, \vect x)
    \end{cases}.
\end{equation}
In (\ref{E:pH}), $\vect x \in \mathbb{C}^n$ is the state vector, $\vect u \in \mathbb{C}^n$ and $\vect y \in \mathbb{C}^m$ are the input and output vectors, %$\vect w \in \mathbb{C}^n$ is a power input whose meaning will become clear later with a concrete system; 
$\matr J(t, \vect x) \in \mathbb{C}^{n\times n}$ is the time-varying interconnection matrix that is skew-Hermitian,
$\matr R \in \mathbb{C}^{n\times n}$ is the constant damping matrix, $\matr G \in \mathbb{C}^{n\times m}$ is the input matrix, $H(t, \vect x) \geq 0$ is the time-varying Hamiltonian, and $\nabla_{\vect x} H(t, \vect x)$ is the complex gradient with respect to the inner product $\langle \vect y, \vect x \rangle = \Re\{\vect y^\herm \vect x \}$.\footnote{The subscript for the complex gradient in $\nabla_{\vect x}H(t, \vect x)$ is omitted in the sequel.} We assume the following:
\begin{enumerate}
    \item[(i)] $\matr J(t, \vect x)$ is full-rank.
    \item[(ii)] $H(t, \vect x)$ is uniformly strictly convex, i.e., for some $a > 0$, it holds that
    \begin{equation} \label{E:strictly_convex}
        D^2 H(t, \vect x) - a \matr I_n \succeq 0,\, \text{for all $t \in \mathbb{R}$},
\end{equation}
    and $H(t, \vect x) = 0 \Leftrightarrow \vect x = 0_n$.
\end{enumerate}
%The Hamiltonian $H(t, \vect x)$ is assumed to be uniformly strictly convex: for some $a > 0$,
%\begin{equation} \label{E:strictly_convex}
    %&\exists\, \vect x'(t),\, H(t, \vect x'(t)) = 0 \text{ and }  H(t, \vect x) > 0 \text{ otherwise} \tag{positive-definiteness} \\
%    D^2 H(t, \vect x) - a \matr I_n \succeq 0,\, \text{for all $t \in \mathbb{R}$}.
%\end{equation}
%and \emph{positive-definite:} there is an $\vect x'(t)$ such that
%\begin{equation} \label{E:positive_definite}
%    H(t, \vect x) = 0 \text{ if $\vect x = \vect x'(t)$ and }  H(t, \vect x) > 0 \text{ otherwise.}
%\end{equation}

For the main result on contraction, we consider a ``closed'' pH system that is without input or output, which is a term coined by J. C. Willems~\cite{willems2007behavioral}. In the port-Hamiltonian model of a network system~\cite{fiaz2013port}, each edge is modeled as an ``open'' pH system, and the input to every edge is mapped from every output by the network constraints:
\begin{equation*}
    \col(\vect u_i) = \matr W \col(\vect y_i)
\end{equation*}
for some skew-Hermitian network matrix $\matr W$. The connected system is written as
\begin{equation} \label{E:closed}
    \dot{\vect x} = (\matr J(t, \vect x) - \matr R) \nabla H(t, \vect x)
\end{equation}
where
\begin{align*}
    &\vect x = \col(\vect x_i),\, H(t, \vect x) = \sum_{i} H_i(t, \vect x_i),\, \matr R = \diag(\matr R_i), \\
    &\matr J(t, \vect x) = \diag(\matr J_i(t, \vect x_i)) + \diag(\matr G_i) \matr W \diag(\matr G_i^\herm).
\end{align*}
The closed pH system (\ref{E:closed}) has an inherent energy balance relation:
\begin{align}
    \dot H(t, \vect x) &= \frac{\partial}{\partial t}H(t, \vect x) + \Re\{\nabla H(t, \vect x)^\herm (\matr J(t, \vect x) - \matr R) \nabla H(t, \vect x) \} \notag \\
    &= \frac{\partial}{\partial t}H(t, \vect x) - \nabla H(t, \vect x)^\herm \matr R \nabla H(t, \vect x), \label{E:energy_balance}
\end{align}
where $\matr J(t, \vect x)$ is energy-preserving because it is skew-adjoint with respect to the assumed inner product.

%\begin{lemma} \label{lem_interconnection}
%Consider a set of pH systems $\Sigma_i,\, i = 1,\ldots,\, N$, each of which verifies $\matr R_i > 0$ and $H_i(t, \vect x_i)$ positive-definite and strictly convex, and a skew-Hermitian network, i.e.,
%\begin{equation*}
%    \col(\vect u_i) = \matr W \col(\vect y_i)
%\end{equation*}
%for a skew-Hermitian $\matr W$. Then the connected system is a pH system with no input satisfying the same assumptions.
%\end{lemma}

%We refer to~\cite{barabanov2019contraction,yaghmaei2023contractive} for existing conditions for contractive pH systems.

To the best of our knowledge, there exist only two groups of papers dedicated to the contraction of pH systems. In~\cite{barabanov2019contraction}, which extends~\cite{yaghmaei2017trajectory}, two LMI and one BMI condition are proposed for contractive pH systems. The conditions require lower and upper bounds on the Hessian $D^2 H(t, \vect x)$, and in~\cite[Prop.~3]{barabanov2019contraction}, the interconnection matrix is required to be bounded relative to the damping matrix, similar to the contraction condition for a general NLTV system. In~\cite{yaghmaei2023contractive}, the partial contraction is used to decouple the state-dependence of $\matr A(\vect x) = \matr J(\vect x) - \matr R(\vect x)$ from the dynamics, which results in a nonlinear matrix inequality contraction condition; it relies critically on the assumption that the Taylor expansion of the matrix-valued function $\matr A(\vect x)$ has no first-order term. Both of the two group of results impose upper bound on $D^2 H(t, \vect x)$, and constraints on the interconnection (skew-adjoint) $\matr J(t, \vect x)$ of the dynamics. The main results in Proposition~\ref{prop_main} to~\ref{prop_convergence} are free of these constraints; the only additional assumption is for the damping matrix $\matr R$ to be a constant. This assumption is satisfied at least by the electromagnetic power system model, as the main application in this paper.

\section{Horizontal Contraction of PH System} \label{sec_contraction}

%The goal is to show that difference in values of the Hamiltonian along any two solutions of (\ref{E:pH}) converges to zero. The basic tool we will use is contraction in quotient space.

The goal is to introduce a special quotient space associated with the pH system (\ref{E:pH}) (the canonical quotient space) and to show that the system is contractive with respect to the quotient distance (horizontal contraction). It is then shown that a direct consequence of this property is that the Hamiltonian of the pH system is convergent.

\subsection{The Canonical Quotient Space}
A quotient space is a partition of the original space $\mathbb{C}^n$ into subsets called equivalence classes. Every point in the same equivalence class is equivalent; that is distance between them is set to zero. The distance between equivalence classes is defined in terms of a Finsler-like distance. Before we give the definition for this distance, we first define the canonical quotient space of the pH system.

For the pH system (\ref{E:closed}), at every time instance $t$, let us consider the quotient space where every equivalence class is an integral curve of the vector field with parameter $t$:
\begin{equation} \label{E:generator_system}
    F_t: \vect x \mapsto \matr J(t, \vect x) \nabla H(t, \vect x).
\end{equation}
The equivalence class of every $\vect x_0 \in \mathbb{C}^n$ at time $t$ is
\begin{equation} \label{E:equivalence_class}
    [\vect x_0]_t = \big\{\Phi_t(\tau, \vect x_0)\mid \tau\in \mathbb{R}\big\}
\end{equation}
where $\Phi_t(\tau, \vect x_0)$ is the integral curve of (\ref{E:generator_system}) from the initial condition $\vect x_0$, i.e.,
\begin{align*}
    \frac{d}{d\tau} \Phi_t(\tau, \vect x_0) &= F_t(\Phi_t(\tau, \vect x_0)) \\
    &= \matr J(t, \Phi_t(\tau, \vect x_0)) \nabla H(t, \Phi_t(\tau, \vect x_0))
\end{align*}
Equivalently, the integral curve $\Phi_t(\tau, \vect x_0)$ is conceptually the solution of the following system with parameter $t$:
\begin{equation} \label{E:generator2}
    \frac{d}{d\tau} \vect x(\tau) = \matr J(t, \vect x(\tau)) \nabla H(t, \vect x(\tau))
\end{equation}
where $\tau$ is the independent variable.
Since $\matr J(t, \vect x)$ is skew-Hermitian, we obtain that the value of $H(t, \vect x)$ is constant at every $\vect x = \Phi_t(\tau, \vect x_0)$ for $\tau \in \mathbb{R}$. This is because (\ref{E:generator2}) entails the energy balance $\frac{d}{d\tau} H(t, \vect x(\tau)) = 0$ where $t$ is a parameter. Hence, at every time instance $t$, the equivalence class $[\vect x_0]_t$ belongs to the level set of $H(t, \vect x)$ for $\vect x_0$; that is,
\begin{equation} \label{E:contain_in_level}
    [\vect x_0]_t \subset \big\{\vect x \in \mathbb{C}^n \mid H(t, \vect x) = H(t, \vect x_0) \big\}. 
\end{equation}
%By (\ref{E:last_assumption}), $\vect x = \eqm{\vect x}(t)$ is the only equilibrium point of (\ref{E:generator_system}). Hence every equivalence class is closed. They are bounded because the level sets of $H(t, \vect x)$ are bounded. Hence every equivalence class is compact.

%Combined with that every solution is contained in a bounded level set, we have that every equivalence class is compact.

Given the definition of the (time-dependent) equivalence classes (\ref{E:equivalence_class}), we can define a (time-dependent) distance measure for any two points $\vect x_1, \vect x_2 \in \mathbb{C}^n$. 
We choose to work with the inner product, 
\begin{equation} \label{E:assumed_inner_product}
    \langle \vect y, \vect x \rangle = \Re\{\vect y^\herm \matr R^{-1} \vect x \}. 
\end{equation}
The definitions of norm and orthogonal subspaces for $\mathbb{C}^n$ are consistent with (\ref{E:assumed_inner_product}). However, with a slight violation of this convention, the definition of the complex gradient $\nabla H(t, \vect x)$ in the pH system (\ref{E:pH}) is not adapted to (\ref{E:assumed_inner_product}). This inconsistency is unimportant because the subsequent contraction analysis concerns the derivative of the RHS of (\ref{E:pH}), i.e., second derivatives of $H(t, \vect x)$. It is chosen to simplify notation.
%However, for familiarity, we do not change the definition of the gradient $\nabla H(t, \vect x)$ or the skew-adjoint part $\matr J(t, \vect x)$ according to (\ref{E:assumed_inner_product}). In other words, we do not change the basic form of the pH system introduced in (\ref{E:pH}), which is based on the inner product $\langle \vect y, \vect x \rangle = \Re\{ \vect y^\herm \vect x \}$. % so that $\nabla H(\vect x)$ is the gradient vector to be used with the original inner product to obtain the time derivative of $H(\vect x)$.

At every time instance $t$, the distance measure according to equivalence classes (quotient distance) is defined as follows. 
By the definition in (\ref{E:generator_system}),
$\big\{\matr J(t, \vect x) \nabla H(t, \vect x)\big\}^\perp$
is the tangent subspace perpendicular to the boundary of the equivalence class at $\vect x$.
For $\vect x \neq 0_n$, define the local projection operator that projects tangent vectors onto this subspace as
\begin{equation} \label{E:projection}
    \mathcal P(t, \vect x) \delta = \delta - \frac{\langle \matr J(t, \vect x) \nabla H(t, \vect x), \delta \rangle}{\| \matr J(t, \vect x) \nabla H(t, \vect x)\| \|\delta\|} \matr J(t, \vect x) \nabla H(t, \vect x),
\end{equation}
where $\delta \in \mathbb{C}^n \neq 0_n$ is the tangent vector of a curve segment to be defined. It is well-defined because the denominator in (\ref{E:projection}) is nonzero by the two assumptions near (\ref{E:strictly_convex}). Then, the quotient distance between two points $\vect x_1, \vect x_2 \in \mathbb{C}^n$ is defined as the integral of the projected infinitesimal curve segment in (\ref{E:projection}) along a minimizing curve from $\vect x_1$ to $\vect x_2$. 
To be precise, consider a piecewise smooth curve $\gamma: [0, 1] \to \mathbb{C}^n$ such that $\gamma(0) = \vect x_1,\, \gamma(1) = \vect x_2$, and $\frac{\partial\gamma}{\partial s}(s) \neq 0$. Denote the set of all such curves as $\Gamma(\vect x_1, \vect x_2)$. The quotient distance is defined as
\begin{equation} \label{E:quotient_distance}
    \dist(t, \vect x_1, \vect x_2) = \inf_{\gamma \in \Gamma(\vect x_1, \vect x_2)} \int_0^1 \bigg\|\mathcal P(t, \gamma(s)) \frac{\partial\gamma}{\partial s}(s)\bigg\|\, ds.
\end{equation}
The existence of a minimizing curve in (\ref{E:quotient_distance}) is guaranteed by the Gauss lemma~\cite[Ch.~6]{bao2012introduction}.
We list the following properties of the quotient distance (\ref{E:quotient_distance}):
\begin{enumerate}
    \item[(i)] $\dist(t, \vect x, \vect y) = \dist(t, \vect y, \vect x)$.
    \item[(ii)] $\dist(t, \vect y, \vect x) \leq \dist(t, \vect y, \vect z) + \dist(t, \vect z, \vect x)$.
    \item[(iii)] $\dist(t, \vect y, \vect x) = 0$ if $[\vect y]_t = [\vect x]_t$ and $\dist(t, \vect y, \vect x) \geq 0$ otherwise.
    \item[(iv)] If $H(t, \vect y) \neq H(t, \vect x)$, then $\dist(t, \vect y, \vect x) > 0$.
\end{enumerate}

The proof is given in the Appendix.

\begin{remark}
The projection $\mathcal P(t, \vect x)$ defines a local $(n {-} 1)$-dimensional tangent subspace.
Contraction in a tangent subspace (or a horizontal distribution) is referred to as horizontal contraction in Section III-A of~\cite{forni2013differential} where the motivation is to not enforce contraction in the symmetry directions. Note, however, that, for the quotient space defined in (\ref{E:generator_system}) and (\ref{E:equivalence_class}), we do not assume that the dynamics of the system preserve the equivalence classes, which is a scenario called a quotient system in Section III-B of~\cite{forni2013differential}.
\end{remark}

\subsection{Horizontal Contraction in the Quotient Space}

The main contraction results are stated without proofs in this subsection. Their proofs are provided in the Appendix.

%Contraction in the quotient space is defined in terms of contraction of the quotient distance.

\begin{definition}
Let $\Phi(t, t_0, \vect x_0)$ be the solution of the pH system (\ref{E:pH}) from the initial condition $(t_0, \vect x_0)$.
The pH system (\ref{E:pH}) is said to be \emph{horizontally contracting in the canonical quotient space} (HC for short) if, for some $c > 0$ (the contraction rate), it holds that, for every $\vect x_1, \vect x_2 \in \mathbb{C}^n$ and $t_0\in \mathbb{R}$,
\begin{equation} \label{E:s1-0} 
    \dist(t, \Phi(t, t_0, \vect x_1), \Phi(t, t_0,  \vect x_2)) \leq e^{-c(t - t_0)} \dist(t_0, \vect x_1, \vect x_2).
\end{equation}
The pH system is said to be weakly HC if (\ref{E:s1-0}) holds with $c = 0$. \hfill $\lozenge$
\end{definition}

%For short, we simply say that a pH system is HC when it is HC in the canonical quotient space.

%The main result is stated as follows.

The following two results concern the intrinsic contraction properties of the pH system.

\begin{proposition}[Horizontal Contraction with $\matr R \succ 0$] \label{prop_main}
%Consider the pH system (\ref{E:pH}) with $\vect u \equiv 0$. 
The closed pH system (\ref{E:closed}) that has a uniformly strictly convex Hamiltonian, i.e., condition (\ref{E:strictly_convex}), 
is HC with the contraction rate, 
\begin{equation*}
    c = \min_{t \in \mathbb{R}} \lambda_{\min}(D^2 H(t, \vect x))\, \lambda_{\min}(\matr R),
\end{equation*}
where $\lambda_{\min}(\matr A)$ is the smallest eigenvalue of the Hermitian matrix $\matr A$. \hfill $\lozenge$
%Then the following relation about contraction in quotient distance holds: for every $\vect x_1, \vect x_2 \in \mathbb{C}^n$,
%\begin{equation} \label{E:s1-0} 
%    \dist(t, \Phi(t, t_0, \vect x_1), \Phi(t, t_0,  \vect x_2)) \leq e^{-c(t - t_0)} \dist(t_0, \vect x_1, \vect x_2),
%\end{equation}
%where $\Phi(t, t_0, \vect x)$ is the solution of the pH system from the initial condition $(t_0, \vect x)$, and $c = \lambda_{\min}(D^2 H(t, \vect x))\, \lambda_{\min}(\matr R)$.
\end{proposition}

\begin{proposition}[Weak Horizontal Contraction with $\matr R = \matr 0$] \label{prop_no_diss}
The closed pH system (\ref{E:closed}) that has a uniformly strictly convex Hamiltonian, i.e., condition (\ref{E:strictly_convex}), and zero dissipation, i.e., $\matr R = \matr 0$, is weakly HC. \hfill $\lozenge$
\end{proposition}

\begin{remark}
The classical Hamiltonian dynamics with a time-varying strictly convex Hamiltonian is covered by Proposition~\ref{prop_no_diss} with the interconnection matrix $\matr J = \big[\matr 0,\, \matr I; -\matr I,\, \matr 0 \big]$. \hfill $\lozenge$
\end{remark}

The following two results concern the implications of HC on the behavior of the solutions.

%\begin{proposition} \label{prop_limit_set}
%Consider the closed and HC pH system (\ref{E:closed}). Assume that the Hamiltonian $H(t, \vect x)$ is bounded along every solution of the system. Then,
%for every $\vect x_0 \in \mathbb{C}^n$ and $t_0 \in \mathbb{R}$,
%$\lim_{t\to \infty} H(t, \Phi(t, t_0, \vect x_0)) = \bar H$
%for some $\bar H \geq 0$. \hfill $\lozenge$
%, the value of $H(\Phi(t, t_0, \vect x_0))$ converges to the value of $H(t, \vect x)$ on the limit cycle as $t \to \infty$.
%\end{proposition}

\begin{proposition}[Converging Hamiltonian Difference] \label{prop_limit_cycle}
Consider the closed and HC pH system (\ref{E:closed}). Let $\eqm{\vect x}(t)$ be a particular solution. Then, the Hamiltonian value converges to $H(t, \bar{\vect x}(t))$ from every initial value, i.e., %$H(t, \bar{\vect x}(t)) = \bar H$ for some $\bar H \geq 0$, and, for every initial condition $\vect x_0$ at $t_0$,
from every initial condition $(t_0, \vect x_0)$, $\lim_{t\to \infty} H(t, \Phi(t, t_0, \vect x_0)) - H(t, \eqm{\vect x}(t)) = 0$.  \hfill $\lozenge$
\end{proposition}

%\begin{remark}
%The symmetry $T$ in Proposition~\ref{prop_symmetry} includes both discrete and continuous symmetry. Examples of discrete symmetry include reflection and permutation. Examples of continuous symmetry include continuous rotation and translation. When the system possesses a symmetry, it is usually true that the Hamiltonian, which represents energy storage, is invariant to the symmetry transformation; hence the second assumption in Proposition~\ref{prop_symmetry} is reasonable.
%\end{remark}

%\begin{corollary} \label{cor_stability}
%Assume that the HC pH system in Proposition~\ref{prop_main} has a limit cycle $\eqm{\vect x}: \mathbb{T} \to \mathbb{C}^n$. Then, from any $(t_0, \vect x_0)$, the value of $H(\Phi(t, t_0, \vect x_0))$ converges to the value of $H(t, \vect x)$ on the limit cycle as $t \to \infty$.
%\end{corollary}

%\begin{remark}
%By the forward contraction argument~\cite{forni2013differential}, it is implied from Corollary~\ref{cor_stability} that the Hamiltonian is constant on a limit set, e.g., the orbit of a limit cycle.
%\end{remark}

\begin{remark}
By definition, every equivalence class of the canonical quotient space is $1$D; meanwhile, every level set of the Hamiltonian $H(t, \vect x)$ for a fixed $t$ has dimension $2n - 1$ (since $\mathbb{C}^n$ has the same dimension as $\mathbb{R}^{2n}$). For $n = 1$ (every level set is $1$D and coincides with an equivalence class), then HC implies difference in the Hamiltonian values converges at the exponential rate $e^{-ct}$. For $n > 1$ (every level set has dimension higher than one), then exponential contraction of the difference in the Hamiltonian values is not guaranteed. \hfill $\lozenge$
%Each equivalence class parameterized by $\tau$ is $1$-D, whereas the level sets of the Hamiltonian are $(2n - 1)$-D. This allows the Hamiltonian of a HC system to temporarily diverge for finite $t$. For this reason, HC does not imply contraction in the Hamiltonian. A special case is $2$-D (planar) systems where the $1$-D equivalence classes coincide with the level sets. In this case, the Hamiltonian is contractive iff the system is HC. 
\end{remark}

The following is the main condition for the convergence of the Hamiltonian, i.e., convergence to a single constant value.

\begin{proposition}[Hamiltonian Convergence Principle] \label{prop_convergence}
Consider the closed and HC pH system (\ref{E:closed}). Assume that the set in which the Hamiltonian has zero derivative, i.e.,
\begin{equation*}
    E_t = \left\{ \vect x \in \mathbb{C}^n \mid \frac{\partial}{\partial t} H(t, \vect x) - \nabla H(t, \vect x)^\herm \matr R \nabla H(t, \vect x) = 0 \right\},
\end{equation*}
is time-independent.
Then, the Hamiltonian value converges from every initial value, i.e., from every initial condition $(t_0, \vect x_0)$, $\lim_{t\to \infty} H(t, \Phi(t, t_0, \vect x_0)) = \eqm H$ for some constant $\eqm H \geq 0$. \hfill $\lozenge$
\end{proposition}

The set $E_t$ in Proposition~\ref{prop_convergence} is usually found to be time-independent because the system can be alternatively written as a time-invariant system. In most applications, including the power system model to be introduced, the time dependence of the Hamiltonian represents a power source; that is,
\begin{equation*}
    \eta(\vect x) = \frac{\partial}{\partial t} H(t, \vect x)
\end{equation*}
represents the the input power.\footnote{We refer the reader to~\cite{krhavc2024port,monshizadeh2019power} for some perspectives on port-Hamiltonian system with power input.} It is a mathematical technique to represent state-dependent power input/output without introducing negative eigenvalues into the damping matrix $\matr R$. If a time-invariant form of the system exists, then the set $E_t$, which represents the states where the stored energy is steady, can be alternatively defined by a time-independent condition, and is hence a time-independent set.
%For these systems, it is usually also true that the dissipation is dependent on the state only so that $E_t$ is time-invariant.

When the system has a power input, the limit set or steady state usually cannot be described by an equilibrium point (for example the van der Pol equation~\cite{Khalil:1173048}). In this case, the limit set is the result of a balance between the power input and dissipation. For studying these systems, Proposition~\ref{prop_convergence} asserts that, if the power input can be mathematically expressed as a time-dependence part of the Hamiltonian, then the limit set is contained in a level set of the original time-independent Hamiltonian. 
On the possible types of limit sets,
if the dimension of the system is $3$ and the dimension of the level set is $2$, then by the Poincare-Bendixson theorem~\cite{Khalil:1173048}, the limit set must be a limit cycle. If the dimension of the system greater than $3$, more complicated limit sets may exist (a chaos).

\section{Stability of the Electromagnetic Power System Model} \label{sec_stability}
The electromagnetic model or the fundamental model is different from the electromechanical model in that the fundamental inductor and capacitor equations are included, rather than simplified into linear impedance equations. The impedance equations are based on the assumption of a steady state of a synchronized frequency throughout the system, omitting all other types of limit cycles and necessitating separate harmonic power flow studies~\cite{xia1982harmonic}.
The difficulty in studying the electromagnetic model is twofold. First, the dimension of the system including the inductor fluxes and the capacitor charges is much higher. Second, the inductor and capacitor dynamics are much faster than the mechanical dynamics of the synchronous generator (SG). Attempts on the hard problem of extending the electromechanical stability conditions to the electromagnetic model lead to overly conservative conditions~\cite{gross2019effect,subotic2020lyapunov}. 

We consider in this section the stability of the electromagnetic power system model. As the main application of Proposition~\ref{prop_convergence}, we show that the convergence of the Hamiltonian is sufficient for stability assuming that a desirable limit cycle steady state exists. To this end, we first introduce the physical model of the SG, followed by its formulation into a time-varying pH system that is covered by Proposition~\ref{prop_convergence}. Then, the model of a power system with two SG is provided as an example. Lastly, the stability of this model is proved.

%We will first model the SG in the stator's $\alpha\beta$-coordinates as a pH system, followed by the model of the power network with constant impedance load. Lastly, Proposition~\ref{prop_convergence} is applied to characterize the stability of a two-machine power system. %The motivation is simple, the model of the SG in the $0dq$-reference frame require an additional angle variable for the rotor's angle. We would want to eliminate this angle variable because it, being defined on the periodic torus space, is known to contradict with the contraction analysis.

\subsection{Physical Model in $\alpha\beta$-Coordinates}
Assume the motor sign convention, i.e., positive stator current goes into the machine.
The mechanical dynamics is given by the swing equation:
\begin{align}
    J \dot{\omega} &= -F \omega - T_e + T_m \label{E:swing1} \\
    \dot\theta &= \omega, \label{E:swing2}
\end{align}
where $\omega$ is the rotor angular frequency, $\theta$ is the rotor angle, $J$ is the rotational inertia, $F$ is the viscous damping, and $T_e$ and $T_m$ are respectively the (accelerating) electrical and mechanical torque.\footnote{We refer the reader to~\cite{forni2014differential,efimov2017relaxed,efimov2019boundedness} for perspectives on the difficulty in studying the stability of systems in periodic angle spaces, i.e., the $2\pi$ periodic torus space $\mathbb{T}$ in which $\theta$ lives.}
%where $J$ is the rotational inertia; $F$ is the viscous damping.
%Note that here $J$ is the rotational inertia rather than the inertia constant $M = J \omega$ that is found in the version of the swing equation assumed in power engineering. The difference is that, in here, (\ref{E:swing1}) is not multiplied through by $\omega$ to get electrical and mechanical power to replace the electrical and mechanical torque. The problem with the modified swing equation is that the inertia constant $M$ is assumed to be a constant, which can result in inaccurate dynamics~\cite{caliskan2015uses}. 

We assume balanced condition such that the $0$-phase in the stationary $\alpha\beta0$-coordinates is a decoupled DC system whose state is constant zero~\cite{o2019geometric}.
Choose the complex variable $I = I^\alpha + j I^\beta$ for the stator current and $V = V^\alpha + j V^\beta$ for the terminal voltage. Let $\psi = M I_F \in \mathbb{R}$ be the constant field flux, i.e., mutual inductance $M$ times the field current $I_F$. The stator equation is given by
%\begin{align}
%    \dot{(LI + \psi e^{j\theta})} = -R I + V, \label{E:stator}
%\end{align}
%which can be written alternatively to reveal the internal EMF in the RHS as follows,
\begin{align*}
    L \dot{I} = -\psi (j\omega e^{j\theta}) - R I + V,
\end{align*}
where $R$ is the stator current, and $-\psi (j\omega e^{j\theta})$ is the internal EMF. %The equation for the shunt capacitor at the SG bus is given by
%\begin{equation*}
%    C \dot{V} = -G V - I + I_b
%\end{equation*}
%where $I_b$ is the bus current.
To complete the swing equation (\ref{E:swing1}), note that the electrical torque is equal to the power transfer divided by the frequency:
\begin{align*}
    T_e &= \frac{\Re\{-\psi (j \omega e^{j\theta}) I^\herm \}}{\omega} = \Re\{ j \psi e^{-j\theta} I\},
\end{align*}
%where we substituted $I e^{-j\theta} = I^r + j I^i$ in the last equality. 
Assume that the mechanical source has droop characteristic:
\begin{equation*}
    T_m = T_0 - F_1 \omega,
\end{equation*}
where $F_1$ is the torque droop ratio, and $T_0$ is the projected zero-freqeuncy torque.
Then we obtain the controlled swing equation,
\begin{equation*}
    J \dot\omega = -F \omega - \Re\{j\psi e^{-j\theta} I\} + T_0,
\end{equation*}
where we absorbed $F \leftarrow F + F_1$.

\subsection{Time-Varying PH Model} 
Choose the state vector $\vect x = \big[ \vect x_1,\, \theta \big]^\tran \in \mathbb{R} \times \mathbb{C} \times \mathbb{T}$ where
\begin{equation} \label{E:state}
    \vect x_1 = \big[ x_1,\, x_2 \big]^\tran = \big[J \omega - T_0 t,\, L I \big]^\tran.
\end{equation}
The space $\mathbb{T}$ is the $2\pi$-periodic torus~\cite{FB-ADL:04} for the angle. 
%The angle symmetry can then be expressed as
%\begin{equation*}
%    \mathcal T(\tau) \vect x = e^{j\tau} \vect x.
%\end{equation*}
The equations for the state vector are given by
\begin{align}
    \frac{d}{dt} (J \omega - T_0 t) &= -F \omega - \Re\{j\psi \omega e^{-j\theta} I\} \label{E:sg_1} \\
    \frac{d}{dt} (L I) &= -j \psi \omega e^{j\theta} - R I + V \\
    %\frac{d}{dt} (C V) &= -G V - I + I_b \\
    \frac{d}{dt} \theta &= \omega. \label{E:sg_2} 
\end{align}
Define the Hamiltonian as
\begin{equation*}
    H(t, \vect x_1) = \frac{1}{2} J^{-1} (x_1 + T_0 t)^2 + \frac{1}{2} L^{-1} \big\|x_2 \big\|^2, % + \frac{1}{2} C^{-1} \big\|x_3\big\|^2. \notag
\end{equation*}
which is equal to the inertial energy plus the magnetic energy.
The gradient of the Hamiltonian is
\begin{align}
    \nabla H(t, \vect x_1) &= \begin{bmatrix}
        J^{-1} (x_1 + T_0 t) \\
        L^{-1} x_2
    \end{bmatrix} = \begin{bmatrix}
        \omega \\
        I
    \end{bmatrix}. \label{E:gradient}
\end{align}
Note that the Hamiltonian can be alternatively expressed as
\begin{equation*}
    H(t, \vect x_1) = H(\vect s) = \frac{1}{2} \vect s^\herm \matr Q^{-1} \vect s
\end{equation*}
with the co-state $\vect s = \nabla H(t, \vect x)$ and $\matr Q = \mathrm{diag}(J^{-1}, L^{-1})$. We can find the real Hessian of the Hamiltonian as\footnote{The set of independent variables are $\{\vect x,\,  t\}$. All partial derivatives are defined with respect to these independent variables.}
\begin{equation} \label{E:sg_hessian}
    D^2 H(t, \vect x_1) = \begin{bmatrix}
        J^{-1} &0 \\
        0 &L^{-1} \matr I
    \end{bmatrix}
\end{equation}
for the equivalent real state vector, $\mathcal U \vect x_1 = \big[ J\omega - T_0 t,\, L I^\alpha,\, L I^\beta \big]^\tran$. Since $J, L > 0$, the Hamiltonian is uniformly strictly convex.
Based on (\ref{E:state}) and (\ref{E:gradient}), the pH model for $\vect x_1$ is obtained as
\begin{equation} \label{E:SM_model}
    \Sigma_{sg}: \begin{cases}
        \dot{\vect x}_1 = \mathcal W \big[ (\matr J(\theta) - \matr R) \nabla H(t, \vect x_1) + \matr G \vect u \big] \\
    \vect y = \matr G^\herm \nabla H(t, \vect x_1)
    \end{cases},
\end{equation}
where $\vect u = V,\, \vect y = I$,
\begin{equation*}
    \mathcal W = \begin{bmatrix}
        \Re &0 \\
        0 &1
    \end{bmatrix},\, \matr G = \begin{bmatrix}
        0 \\
        1
    \end{bmatrix},
\end{equation*}
and 
\begin{align*}
    &\matr J(\theta) = \begin{bmatrix}
        0 &-j \psi e^{-j\theta} \\
        -j \psi e^{j\theta} &0
    \end{bmatrix},\, \matr R = \begin{bmatrix}
        F &0 \\
        0 &R
    \end{bmatrix}.
\end{align*}
For the dynamics (\ref{E:SM_model}), the inner product assumed is
\begin{equation*}
    \langle \vect y, \vect x \rangle = \Re\{ (\mathcal W \vect y)^\herm (\mathcal W \vect x) \}. 
\end{equation*}

Note that the SG system (\ref{E:sg_1})--(\ref{E:sg_2}) is not exactly an open pH system (\ref{E:pH}); the pH system for the SG in (\ref{E:SM_model}) does include the last equation (\ref{E:sg_2}) while the physical Hamiltonian $H(t, \vect x_1)$ is not dependent on the angle $\theta$. Note, however, that the main propositions in Section~\ref{sec_contraction} still apply to this system with the minimal modification as follows. For the quotient distance in (\ref{E:quotient_distance}), the projection of the $\vect x_1$ dimensions is defined as in (\ref{E:projection}), i.e.,
\begin{equation*}
    \mathcal P(t, \vect x) \begin{bmatrix}
        \delta_1 \\
        0
    \end{bmatrix} = \begin{bmatrix} 
        \delta_1 - \frac{\langle \matr J(\theta) \nabla H(t, \vect x_1), \delta_1 \rangle}{\| \matr J(\theta) \nabla H(t, \vect x_1)\| \|\delta_1\|} \matr J(\theta) \nabla H(t, \vect x_1) \\
        0
    \end{bmatrix}
\end{equation*}
for $\delta_1 \in T_{\vect x_1} (\mathbb{R} \times \mathbb{C}) = \mathbb{R} \times \mathbb{C}$, and the projection of the $\theta$ dimension is set to $0$, i.e.,
\begin{equation*}
    \mathcal P(t, \vect x) \begin{bmatrix}
        0 \\
        0 \\
        \delta \theta
    \end{bmatrix} = \begin{bmatrix}
        0 \\
        0 \\
        0
    \end{bmatrix}
\end{equation*}
for $\delta \theta \in T_{\vect x} \mathbb{T} = \mathbb{R}$.
Then, it is easy to checked that the proofs of the main propositions in Section~\ref{sec_contraction} still work under the respective conditions.

\begin{comment}
Note that, compared to the system considered in Section~\ref{sec_contraction}, the SG system has the additional state variable $\theta$ which does not affect the definition of the Hamiltonian, i.e., $\frac{\partial H}{\partial \theta} = 0$. However, it can be easily checked that all the machinery from Section~\ref{sec_contraction} still works for the system (\ref{E:SM_model}) if we (i) use the equations for $\vect x$ to generate the equivalence classes, i.e., using
\begin{align*}
    \begin{bmatrix}[1.2]
        \dot{\vect x}_1(\tau) \\
        \dot\theta(\tau)
    \end{bmatrix} &= \begin{bmatrix}
        \mathcal W \left[ (\matr J(\vect x(\tau)) - \matr R) \nabla H(t, \vect x_1(\tau)) + \matr G \vect u(\tau)\right] \\
        \omega(\tau)
    \end{bmatrix} \\
    \vect y(\tau) &= \matr G^\herm \nabla H(t, \vect x_1(\tau)),
\end{align*}
where $\vect u(\tau)$ is supplied from the other open subsystems,
and (ii) only include $\vect x_1$ in the definition of the quotient distance in (\ref{E:projection}), i.e.,
\begin{equation*}
    \mathcal P(t, \vect x) \begin{bmatrix}
        0 \\
        0 \\
        1
    \end{bmatrix} = 0_3.
\end{equation*}
In particular, the system (\ref{E:SM_model}) without considering the input $\vect u$, is HC if $\matr R > 0$ and $D^2 H(t, \vect x_1) \geq c > 0$, which are true. The input is supplied from other subsystems of the power system, which are introduced in the next subsection.
\end{comment}

We can check the condition of Proposition~\ref{prop_convergence} as follows, Note that, from (\ref{E:energy_balance}), the power input is
$\frac{\partial}{\partial t} H(t, \vect x_1) = T_0 \omega$,
and the dissipation is
$-\vect s^\herm \matr R \vect s$.
Both are time-independent. Then, by Proposition~\ref{prop_convergence}, the Hamiltonian is convergent if we ignore the input. The input is supplied from other subsystems of the power system, which are introduced in the next subsection.

\begin{comment}
The equivalence classes at time $t$ are defined in the full state space $\vect x$, i.e., from the system
\begin{align*}
    \begin{bmatrix}[1.2]
        \dot{\vect x}_1(\tau) \\
        \dot\theta(\tau)
    \end{bmatrix} &= \begin{bmatrix}
        \mathcal W \left[ (\matr J(\vect x(\tau)) - \matr R) \nabla H(t, \vect x_1(\tau)) + \matr G \vect u(\tau)\right] \\
        \omega(\tau)
    \end{bmatrix} \\
    \vect y(\tau) &= \matr G^\herm \nabla H(t, \vect x_1(\tau)),
\end{align*}
where $\vect u(\tau)$ is supplied from the other open subsystems.
It can be checked that, ignoring the input, Proposition~\ref{prop_convergence} holds by treating $\theta$ similar to $\vect x_1$ except that $\theta$ does not contribute to the distance. %The inner product assumed for checking HC is $\langle \mathcal U \vect y, \mathcal U \matr R^{-1} \vect x \rangle$.
\end{comment}

\begin{remark}
The technique for modeling the SG with constant field current as a pH system can be easily applied to the full-order SG dynamics with one excitation winding and three damper windings~\cite{vittal2019power}. To do this, the DC circuits are modeled as real variables similar to $x_1$ in the above. We choose to present the simpler SG model in this paper to show the idea more clearly. \hfill $\lozenge$
\end{remark}

%\subsection{Single-Machine Infinite-Bus System}

\subsection{Two-Machine System with Constant Impedance Loads}
\begin{figure}[!t]
\subfloat{\includegraphics[width=2.2in,center,margin=0in 0in 0in 0in]{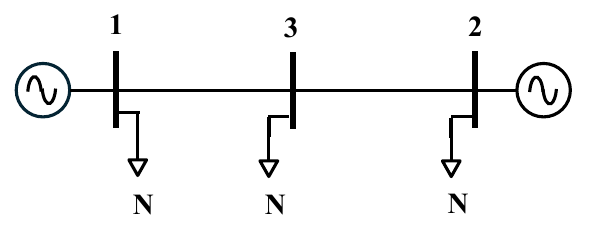}}
\hfill
\subfloat{\includegraphics[width=1.4in,center,margin=0in 0in 0in 0in]{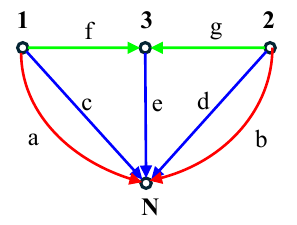}}
\caption{Single-line diagram of the two-machine system and the underlying graph topology (SG: red, shunt capacitor: blue, R--L line: green)}
\label{fig_topology}
\hfill\end{figure}

As an example, consider a two-machine system with constant impedance loads. With the pH modeling technique introduced in~\cite{fiaz2013port}, the system is seen as a directed graph where each edge is either a SG, a shunt capacitor, or an R--L line. The sign convention for the edge voltage and current follows the direction of the edge; that is positive edge voltage and current consumes real power. See Fig.~\ref{fig_topology} for the graph topology of the two-machine system considered.

The two SG systems are denoted as $\Sigma_{sg_i},\, i \in \{1, 2\}$. The transmission lines are modeled by the lumped-parameter $\Pi$-model~\cite{watson2021scalable}. The equations for the shunt capacitor edge is, for $i \in \{3, 4, 5\}$,
\begin{equation} \label{E:capacitor}
    \Sigma_{sh_i}: \begin{cases}
        \frac{d}{dt} (C_i V_i) = -Y_i \nabla H_i(C_i V_i) + I_i \\
        V_i = \nabla H_i(C_i V_i)
    \end{cases},
\end{equation}
where $V_i$ and $I_i$ are respectively the edge voltage and current,
$H_i(\vect x_i) = \frac{1}{2} C_i^{-1} \|\vect x_i\|^2$ with $\vect x_i = C_i V_i$, $Y_i$ with $\Re\{Y_i\} > 0$ is the admittance of the constant impedance load. The equations for the line edges are, for $i \in \{6, 7\}$,
\begin{equation} \label{E:inductor}
    \Sigma_{ln_i} \begin{cases}
        \frac{d}{dt} (L_i I_i) = -R_i \nabla H_i(L_i I_i) + V_i \\
        I_i = \nabla H_i(L_i I_i)
    \end{cases},
\end{equation}
where $V_i$ and $I_i$ are respectively the edge voltage and current; 
$H_i(\vect x_i) = \frac{1}{2} L_i^{-1} \|\vect x_i\|^2$ with $\vect x_i = L_i I_i$; $R_i$ is the series line resistance.

Based on the inputs and outputs,
\begin{align*}
    \vect u &= \big[ V_1,\, V_2,\, I_3,\, I_4,\, I_5,\, V_6,\, V_7 \big]^\tran, \\
    \vect y &= \big[ I_1,\, I_2,\, V_3,\, V_4,\, V_5,\, I_6,\, I_7 \big]^\tran.
\end{align*}
and KCL and KVL, the network matrix which relates the inputs and outputs of the edges, is found as
\begin{equation*}
    \matr W = \begin{bmatrix}
        0 &0 &1 &0 &0 &0 &0 \\
        0 &0 &0 &1 &0 &0 &0 \\
        -1 &0 &0 &0 &0 &-1 &0 \\
        0 &-1 &0 &0 &0 &0 &-1 \\
        0 &0 &0 &0 &0 &1 &1 \\
        0 &0 &1 &0 &-1 &0 &0 \\
        0 &0 &0 &1 &-1 &0 &0
    \end{bmatrix}.
\end{equation*}
We verify that that $\matr W$ is skew-symmetric. 
%For derivation of the interconnection matrix for a power system with a general topology, we refer to~\cite{jiang2024reference}. 
Hence, recalling (\ref{E:closed}), the two-machine system connected through $\matr W$ is a pH system with $\matr R \succ 0$ and $D^2 H(t, \vect x) - a\matr I_n \succeq 0$. Note that the skew-symmetry of the network matrix results from KVL and KCL, and so it holds regardless of the topology of the power system~\cite{jiang2024reference}.

\subsection{Stability of the Two-Machine System}

To prove the stability of the two-machine system, we need the following result on the uniqueness of the limit cycle of an RLC circuit with sinusoidal forcing.

\begin{lemma} \label{lem_RLC}
Consider an RLC circuit with sinusoidal forcing. With the edge dynamics given by (\ref{E:capacitor}) and (\ref{E:inductor}), the equations can be written in the form,
\begin{equation*}
    \dot{\vect x} = (\matr J - \matr R) \nabla H(\vect x) + \vect g u
\end{equation*}
with $u = e^{j\omega_0 t}$ and $H(\vect x) = \frac{1}{2} \vect x^\herm \matr Q \vect x$ for diagonal $\matr Q$ and $\matr R$. Assume the system has a limit cycle $\eqm{\vect x}(t) = e^{j\omega_0 \tau} \eqm{\vect x}(0)$. Then the orbit of the limit cycle is the only possible limit set. \hfill $\lozenge$
\end{lemma}

The proof is given in the Appendix.

%The main result of this paper is stated as follows.

\begin{proposition} \label{prop_final}
Assume the two-machine system has a limit cycle solution $\eqm{\vect x}(t)$ of a synchronized frequency $\eqm\omega \in \mathbb{R}$ such that the complex variables change as $x_i(t) = e^{j\eqm \omega t} x_i(0)$. Then every solution of the system converges to the limit cycle solution as $t \to \infty$. \hfill $\lozenge$
\end{proposition}

The proof is given in the Appendix.

Since the proof for the two-machine system does not rely on the graph topology, the same stability result can be generalized to multi-machine systems with constant impedance loads. For the sake of the analysis, the only difference is the dimensionality of the network matrix. In general, one can consider distributed-parameter model of the transmission lines where the number of edges approaches infinity~\cite{watson2021scalable}.

\begin{proposition}
Assume that a multi-machine system consisting of SG, shunt capacitor, and R--L line edges, has a limit cycle solution of a synchronized frequency. Then every solution of the system converges to the limit cycle solution as $t \to \infty$. \hfill $\lozenge$
\end{proposition}

The proof is similar to the proof of Proposition~\ref{prop_final} and is omitted.

\begin{remark}
By Proposition~\ref{prop_main}, the convergence rate of the Hamiltonian is proportional to $\lambda_{\min}(D^2 H(t, \vect x))\, \lambda_{\min}(\matr R)$. Assuming that the mechanical energy storage is dominant in $H(t, \vect x)$ and the electrical energy dissipation is dominant in $\matr R$, we obtain that the convergence rate of the power system Hamiltonian is estimated as $\min\{\Re\, Y_i\}/\max\{J_i\}$. \hfill $\lozenge$
\end{remark}
\section{Numerical Examples} \label{sec_test}

In Proposition~\ref{prop_final}, we proved that a limit cycle of the power system with constant impedance loads, is globally convergent if it exists. Note that stability in power system traditionally refers the steady-state behavior of the system; this includes
\begin{enumerate}
    \item[1.] convergence of all the SG frequencies to a single constant synchronized frequency,
    \item[2.] no low-frequency oscillations, i.e., the envelope of every three-phase AC signal is a straight horizontal line,
    \item[3.] no harmonics, i.e., every AC signal has a single Fourier component at the synchronized frequency from (i).
\end{enumerate}
We will show that these steady-state instability conditions do not contradict with the global convergence result proved in Proposition~\ref{prop_final} because the traditional stability concerns the regularity of the steady state as opposed to its convergence property. Here we should adopt a geometric approach to the system by considering its set of solutions that has ran for all finite time. The steady state, be it a \emph{synchronized limit cycle} or an \emph{imperfect (negation of 1--3) limit cycle}, is the target set whose Hamiltonian value is convergent, and, in the case of a synchronized limit cycle, the orbit is convergent as well.
In fact, we will show that the instability of the steady-state in traditional power engineering correspond to the nonexistence of a synchronized limit cycle, due to the conflict between the power flow constraint from the RLC network and the power injection constraint from the SGs.

The two machine system in Fig.~\ref{fig_topology} is chosen. A electromagnetic model is build in Simulink Specialized Power System in SI units. The PMSM model is used to model a SG with constant field current. The parameters of the SG are obtained from Example~4.1 from~\cite{vittal2019power}, which are given in Table~\ref{table_parameters} as well as all the other parameters. %We restrict the system to a balanced condition by setting the parameters to be the same in all three phases and restricting the initial condition to be balanced. 
%The code for this test has been made available at~\cite{code}.
%We have conducted the following tests (all in small figures):

\begin{table}[!t]
\renewcommand{\arraystretch}{1.2}
\caption{Parameters of the Two-Machine Test System}
\label{table_parameters}
\centering\begin{tabular}{c|c}
\hline
Description &Parameter \\
\hline
SG mechanical &$J = 2.846\times 10^4$ \unit{kg.m^2}, $F = 85.5601$ \unit{N.m.s}, \\
&$p = 4$, $T_0 = 1\times 10^4$ \unit{N.m} \\
\hline
SG electrical &$R_s = 1.542$ \unit{\milli \Omega}, $L_s = 6.341$ \unit{\milli H}, \\
&$\psi = 39.7877$ \unit{V.s} \\
\hline
Shunt capacitor &$C_{sh,3} = 50 $ \unit{\milli F}, $C_{sh,4} = 100 $ \unit{\milli F}, \\
&$C_{sh,5} = 50 $ \unit{\milli F} \\
\hline
Load &$R_{ld,3} = 1$ \unit{k \Omega}, $L_{ld,3} = 10$ \unit{H} \\
&$R_{ld,4} = 4$ \unit{\Omega}, $L_{ld,4} = 1$ \unit{H} \\
&$R_{ld,5} = 1$ \unit{k \Omega}, $L_{ld,5} = 10$ \unit{H} \\
\hline
R--L line &$R_{ln,6} = 3$ \unit{\Omega}, $L_{ln,6} = 1.061$ \unit{H}, \\
&$R_{ln,7} = 3$ \unit{\Omega}, $L_{ln,7} = 1.061$ \unit{H} \\
\hline
%&$R_{ln,7} = 5$ \unit{Ohm}, $L_{ln,7} = 2.653$ \unit{H} \\
\end{tabular}
\end{table}

\subsection{Global Convergence in the Regular Case}
%plot of the Hamiltonian from many random initial conditions
%plot of zero start including voltage, stored energy of every component

\begin{figure}[!t]
\centering 
\subfloat{\centering \includegraphics[width=1.65in]{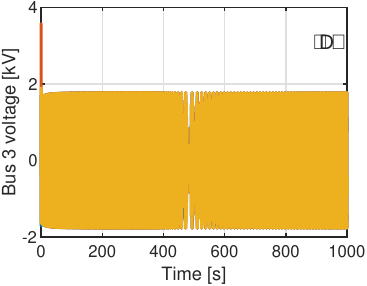}}
\subfloat{\centering \includegraphics[width=1.67in,margin=0.03in 0in]{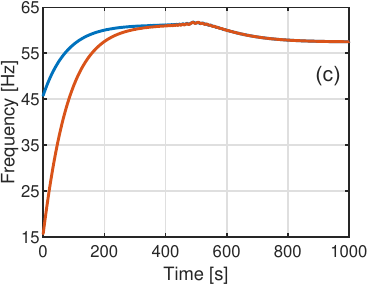}}
\hfil
\subfloat{\centering \includegraphics[width=1.6in,margin=-0.02in -0.04in]{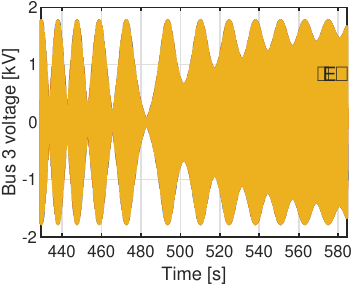}}\hspace{0.03in}
\subfloat{\centering \includegraphics[width=1.69in,margin=0in -0.05in]{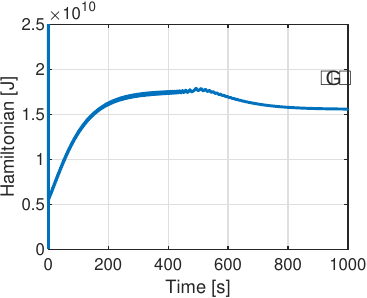}}
\caption{Case where there exists a synchronized limit cycle}
\label{fig_regular}
\hfill\end{figure}

In the first case, we set the parameters of the system such that there is reflection symmetry between the left and right half of the topology. This ensures that a synchronized limit cycle exists so that we can test convergence alone. More generally, a symmetric radial network is similar to a single SG system, and so a synchronized limit cycle exists~\cite{johnson2014synchronization}. We test the convergence property of the system by starting from a random initial condition generated by 100*rand(26) for $26$ real state variables. We can see from Fig.~\ref{fig_regular}d that immediately after the initial condition there is a large overshoot in the Hamiltonian caused by the RLC network quickly returning to an almost quasi steady state. Between $0$~\unit{s} and $485$~\unit{s}, the waveform of the voltage is not regular at all as it exhibits low-frequency oscillation. During this time, the voltage waveform appears constant except for the increasing widths of the distinct wave packets, while the rotor frequencies continue to approach each other. Between $485$~\unit{s} and $700$~\unit{s}, the rotor frequencies are locked to each other, and the low-frequency oscillation is dying. At around $700$~\unit{s}, the system returns to the synchronized limit cycle with a single-frequency waveform. We tested $50$ other initial conditions from 1000*rand(26) to verify that the system returns to the same synchronized limit cycle. We tested multiplying the damping matrix by a factor and observe that the transient time reduces by the same factor. We also tested multiplying the SG inertia by a factor of the nominal value and observed that the qualitative behavior of the system remains the same except that the transient time is extended with higher inertias and shortened with lower inertias. %The reader is invited to verify these observations for themselves with the supplied code~\cite{code}. 

Note that the initial conditions tested here are much farther from the synchronized limit cycle compared to those considered in traditional power system stability studies---the initial condition is far outside the region of convergence predicted by the direct method. Moreover, in traditional power engineering, the excitation control is considered to have a the biggest effect on (steady-state) stability, whereas, here, the field current is kept constant. The transient condition tested here corresponds to uncontrolled black starts or protection device malfunctioning in faults where the rotor angle stability~\cite{tziouvaras2004out} is lost. The test result shows that the traditional stability concepts such as critical clearing time, and the negative effect of low inertia on stability, do not hold at least for this test system. We suspect the reason for these traditional concepts is that it takes two distinct stages for the state to converge. From Fig.~\ref{fig_regular}, there is almost no sign that the low-frequency oscillation is dying before $485$~\unit{s} while the rotor frequencies are approaching each other. This behavior is qualitatively different from the more familiar linear dynamics where convergence is exponential in every state variable. %For the nonlinear power system dynamics, the convergence rate, which is proportional to the damping and the convexity of the Hamiltonian by Proposition~\ref{prop_main}, is only relevant when close to the target limit cycle.

\subsection{Loss of Synchronized Limit Cycle Steady State}
%How Things Can Go Wrong: Zero Torque Droop with High Field Flux}
%plot of 5.5x the base field flux ($\geq 20$ second simulation time) showing the system going into chaotic behavior

%plot of 5x the base field flux showing that the system going into a limit cycle where the frequencies oscillate, and the electrical quantities have undamped harmonics

%plot of 2x the base field flux showing that the system going into a limit cycle where the frequencies converge, and the electrical quantities have undamped harmonics

%In the second case, we increased the input torque of the SG~2 from $1\times 10^4$~\unit{N.m} to $1.5\times 10^4$~\unit{N.m} while keeping the input torque of the SG~1 the same.

\begin{figure}[!t]
\centering 
\subfloat{\centering \includegraphics[width=1.65in]{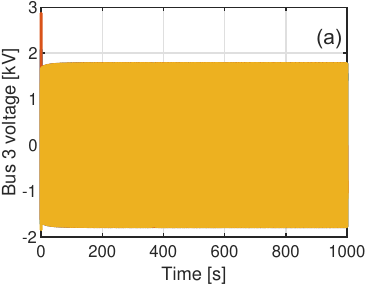}}
\subfloat{\centering \includegraphics[width=1.68in,margin=-0.01in 0in]{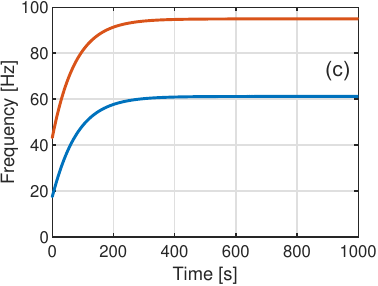}}
\hfil
\subfloat{\centering \includegraphics[width=1.67in,margin=0.02in -0.02in]{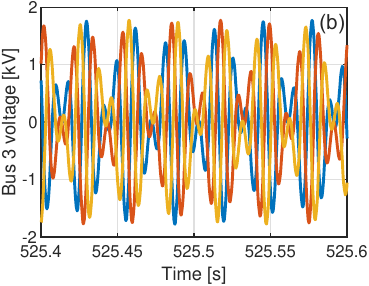}}\hspace{-0.04in}
\subfloat{\centering \includegraphics[width=1.63in,margin=0in -0.02in]{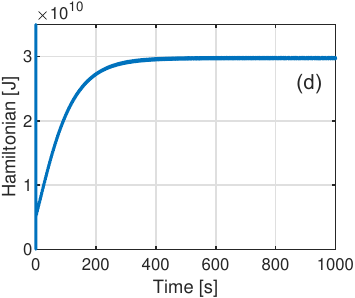}}
\caption{Case where there only exists an imperfect limit cycle with low-frequency oscillation (existence of two steady-state frequencies) due to the large difference between the input torques of the two SGs}
\label{fig_case1}
\hfill\end{figure}

There are several ways that a synchronized limit cycle cannot be reached asymptotically. In the first case, we increased the torque input of SG~2 from $1\times 10^4$~\unit{N.m} to $1.5\times 10^4$~\unit{N.m} so that right side of the two-machine system will inject $1.5$ times the power than the left side if the two rotor frequencies are to converge to the same value. As we can see from Fig.~\ref{fig_case1}, the frequencies of the two machines converge to different values. The frequency of SG~2 is approximately $1.5$ times the frequency of SG~1, possibly due to the additional torque input. The steady state shown in Fig.~\ref{fig_case1}b shows that the steady-state limit cycle exhibits low-frequency oscillation which forms distinct wave packets. This otherwise undamped low-frequency oscillation is dampened in practice by excitation control which modulates the excitation voltage based on measurement of the terminal voltage~\cite{vittal2019power}.
%In the third case, we increased the field flux of both SGs by $2.5$ times.

\begin{figure}[!t]
\centering 
\subfloat{\centering \includegraphics[width=1.67in,margin=0.02in 0in]{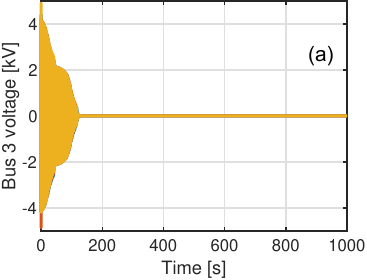}}
\subfloat{\centering \includegraphics[width=1.67in,margin=0in 0in]{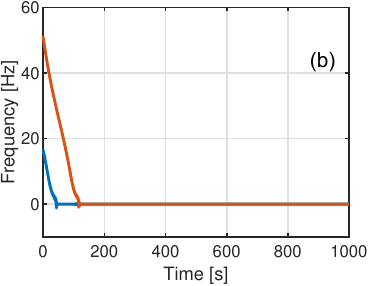}}
\caption{Case where the system collapses to close to zero due to the field fluxes of the SGs being too high}
\label{fig_case2}
\hfill\end{figure}

In the second case, we multiplied the constant field flux of both SGs by $2.5$ times. We can see from Fig.~\ref{fig_case2} that the voltage and the frequency both collapse to close to zero after a transient period. This is because the high field flux results in high EMF, which results in high real power consumption from the constant impedance loads, compared to the relatively low torque input. Note that, even in this case, the final steady state is not exactly zero, but is an equilibrium point that is very close to zero---the zero state is not stable whenever the input torque is nonzero.

In the third case, we observed during our test that, under certain parameter choices, the system goes into a bounded chaotic state rather than a limit cycle. 
%Thirdly, we note that chaotic limiting behavior is possible. However, from the proof of Proposition~\ref{prop_final}, the value of the Hamiltonian of every edge converges independently. This means that all real quantities including the rotor frequencies and the voltage and current amplitudes converge, and the periodic quantities including the voltage and current angles are possibly chaotic.

\subsection{Discussion}

It is shown that global stability, i.e., convergence of every solution to a synchronized limit cycle, holds for multi-machine power system dynamics if the synchronized limit cycle exists. The convergence rate is observed to be proportional to $1/\max\{J_i\}$ and $\min\{\Re\{Y_i\}\}$. It is justified both theoretically and experimentally that the remaining question to ask about power system dynamics is whether the synchronized limit cycle exists, because it is stable if it exists. This explains why the focus on phasor analysis in traditional power engineering has proved successful thus far, despite the highly nonlinear dynamics. The revealed stability property of the traditional SG power system is insightful for the control of inverter-based resources (IBR).
Although the SG dynamics implies globally stability, there is still room to design IBR control schemes to improve transient smoothness and power sharing as the future power system is seeing more frequent reconfigurations. See our work toward this direction in~\cite{jiang2024reference}. %From this perspective, steady-state analysis is still expected to be the most useful tool for power engineers.

From the test results, we can see that the biggest challenge in operating AC power system versus DC power system is not in stability but the potentially complicated steady-state behavior. %There is no fundamental differences between a dissipative AC system and DC system\footnote{Here we omitted the effect of nonlinear loads}---a DC system can be approximated by setting the SG inertia to infinity so that the SG frequency is constrained to $0$. 
The steady state behaviors of AC system include undesirable oscillations and chaos. These important features of the steady state cannot be represented in the traditional phasor analysis, and not fully by harmonic power flow studies. For future research, more research effort is needed toward the precise steady state behaviors of AC power systems to obtain conditions on the existence of synchronized limit cycles and the corresponding steady state control schemes. A notable recent work toward this direction is~\cite{gross2018steady}.
\section{Conclusion} \label{sec_conclusion}

This paper presents a theoretical analysis of the novel CQS property of the time-varying pH system and applied it to the stability analysis of the electronmagnetic model of the power system. It is found that, if the system has a synchronized limit cycle, then its orbit is globally convergent. In the case study of a two-machine system, we verified this stability result and identified that several instability concepts in traditional power engineering are related to the nonexistence of the synchronized limit cycle. The contributions of this paper is threefold. Firstly, it provides a rigorous analysis of power system stability, which unifies the common instability phenomena in power systems. It is identified that the main challenge in the operation of future AC power systems with high power electronics penetration is the existence of a synchronized limit cycle.
%Secondly, it identifies the main challenge in the operation of the future power system as the nonexistence of a synchronized limit cycle. 
Secondly, the converging Hamiltonian principle provides an elegant way to characterize the stability of systems with power sources, exhibiting limit cycle behavior. 
Thirdly, the horizontal contraction property of time-varying pH system is applicable to the stabilization of periodic motions in other areas such as robotics and multi-agent systems.
\appendix
%\section*{Appendix}
%\renewcommand{\thesubsection}{\Alph{subsection}}
%\section{Appendix }

%%% For Appendix A.
% format the equation environment
%\renewcommand{\theequation}{A\arabic{equation}}

% reset the counter
%\setcounter{equation}{0}

\subsection{Proof of Properties (i)--(iv) of the Quotient Distance (\ref{E:quotient_distance})}

%A proof sketch is provided as follows.
\noindent (i) Denote $V(\vect x, \delta) = \|\mathcal P(t, \vect x) \delta \|$. 
Since $V(\vect x, -\delta) = V(\vect x, \delta)$, the pseudo-metric (\ref{E:quotient_distance}) is symmetric, which implies (i). 

\noindent (ii) It is implied from the principle of dynamic programming. 

\noindent (iii) The quotient distance (\ref{E:quotient_distance}) being nonnegative is obvious. If $[\vect x_1]_t = [\vect x_2]_t$, then, by the definition (\ref{E:generator_system}), there is an integral curve of (\ref{E:generator_system}) that joins $\vect x_1$ and $\vect x_2$. The tangent vectors of this integral curve are orthogonal to span of the projection $\mathcal P(\vect x, \delta)$ by definition. Therefore, this curve causes the integral in the RHS of (\ref{E:quotient_distance}) to evaluate to zero, which, combined with nonnegativity, implies $\mathrm{dist}(\vect x_1, \vect x_2) = 0$.

\noindent (iv) From (\ref{E:contain_in_level}), each equivalence class is contained in a level set of $H(t, \vect x)$. Since tangent vector of the equivalence class is $\mathrm{span}(\mathcal P(t, \vect x))^\perp$, $\nabla H(t, \vect x) \in \mathrm{span}(\mathcal P(t, \vect x))$. Then, for each $\vect x, \vect y$ with $H(t, \vect x) \neq H(t, \vect y)$, it holds that $\mathrm{dist}(t, \vect y, \vect x) \geq \min\big\{\|\vect z_1 - \vect z_2\|\mid H(t, \vect z_1) = H(t, \vect x),\, H(t, \vect z_2) = H(t, \vect y)\big\} > 0$, where we expanded each equivalence relation to entire level set for the first inequality and used that level sets are disjoint and closed for the second inequality. \hfill $\square$

%(iii) can be proved by noting that, if $[\vect x_1]_t = [\vect x_2]_t$, the integral curve $\Phi_t(\tau, \vect x_1)$ between $\vect x_1$ and $\vect x_2$ has zero projected tangent vectors everywhere. 
%The ($\Rightarrow$) part of property (iii) is proved by noting that, based on (\ref{E:last_assumption}), every $\Phi_t(\tau, \vect x)$ spans a disjoint compact set.
%(iv) can be proved by noting that, because the equivalence classes are contained in level sets of $H(t, \vect x)$, the tangent subspace $\big\{ \matr J(t, \vect x) \nabla H(t, \vect x) \big\}^\perp$ always contains $\nabla H(t, \vect x)$. Since every curve connecting different level sets has tangent component  in $\nabla H(t, \vect x)$, the quotient distance between two points in different level sets must be strictly positive.

\subsection{Proof of Proposition~\ref{prop_main}}

\noindent \textbf{1.} From Gr\"onwall's lemma, to prove (\ref{E:s1-0}), it suffices to prove that
\begin{align}
    &\frac{d}{d t} \dist(t, \Phi(t, t_0, \vect x_1), \Phi(t, t_0, \vect x_2)) \notag \\
    &\overset{?}{\leq} -c \dist(t, \Phi(t, t_0, \vect x_1), \Phi(t, t_0, \vect x_2)). \label{E:s1-1}
\end{align}
%Based on the chain rule for total derivatives, we separate the contributions to the LHS of (\ref{E:s1-1}) from the two systems,
%\begin{align*}
%    &\dot{\vect x} = \matr J(t, \vect x) \nabla H(t, \vect x) \tag{system 1} \\
%    &\dot{\vect x} = -\matr R \nabla H(t, \vect x). \tag{system 2}
%\end{align*}
%It is clear that system~1 preserves the distance of the quotient space, and so for checking (\ref{E:s1-1}) we only need to consider the contribution from system~2.
Using (\ref{E:quotient_distance}), we can express (\ref{E:s1-1}) as
\begin{align}
    &\frac{d}{d t} \dist(t, \Phi(t, t_0, \vect x_1), \Phi(t, t_0, \vect x_2)) \notag \\
    &= \frac{d}{d t} \min_{\gamma \in \Gamma(\Phi(t, t_0, \vect x_1), \Phi(t, t_0, \vect x_2))} \int_0^1 \bigg\|\mathcal P(t, \gamma(s)) \frac{\partial\gamma}{\partial s}(s) \bigg\|\, ds \notag \\
    &\leq \frac{d}{d t} \int_0^1 \bigg\|\mathcal P(t, \psi(t, s)) \frac{\partial\psi}{\partial s}(t, s) \bigg\|\, ds \label{E:s1-7} \\
    &\overset{?}{\leq} -c \int_0^1 \bigg\|\mathcal P(t, \psi(t, s)) \frac{\partial\psi}{\partial s}(t, s) \bigg\|\, ds \label{E:s1-6} \\
    &= {-}c \dist(t, \Phi(t, t_0, \vect x_1), \Phi(t, t_0, \vect x_2)), \notag 
\end{align}
where $\psi(t, s)$ is a curve that achieves the minimum at $t = t_0$, and $\psi(t, s) = \Phi(t, t_0, \psi(t, s))$ is the solution of the system rooted at $\psi(t_0, s) = \psi(t, s),\, s\in [0, 1]$. The inequality (\ref{E:s1-7}) holds because the curve $\psi(t, s)$ is either minimizing or not for $t > t_0$. Note that, to prove the inequality (\ref{E:s1-6}), it suffices to prove that the integrand satisfies the inequality uniformly, i.e.,
\begin{equation} \label{E:s1-8}
    \frac{d}{d t} \bigg\|\mathcal P(t, \psi(t, s)) \frac{\partial\psi}{\partial s}(t, s) \bigg\|\ \overset{?}{\leq} -c\, \bigg\|\mathcal P(t, \psi(t, s)) \frac{\partial\psi}{\partial s}(t, s) \bigg\|.
\end{equation}
The inequality (\ref{E:s1-8}) is equivalent to
\begin{align}
    &\frac{1}{2} \frac{d}{d t} \Big\langle \mathcal P(t, \psi(t, s)) \frac{\partial\psi}{\partial s}(t, s), \mathcal P(t, \psi(t, s)) \frac{\partial\psi}{\partial s}(t, s) \Big\rangle \notag \\
    &\overset{?}{\leq} -c \Big\langle \mathcal P(t, \psi(t, s)) \frac{\partial\psi}{\partial s}(t, s), \mathcal P(t, \psi(t, s)) \frac{\partial\psi}{\partial s}(t, s) \Big\rangle, \label{E:s1-2}
\end{align}
because, for any $m(t) \geq 0$, $\frac{d}{d t} m \leq -c m \Leftrightarrow \frac{1}{2} \frac{d}{d t} m^2 \leq -c m^2$.

\begin{figure}[!t]
\subfloat{\includegraphics[width=2.5in,center,margin=0in 0in 0in 0in]{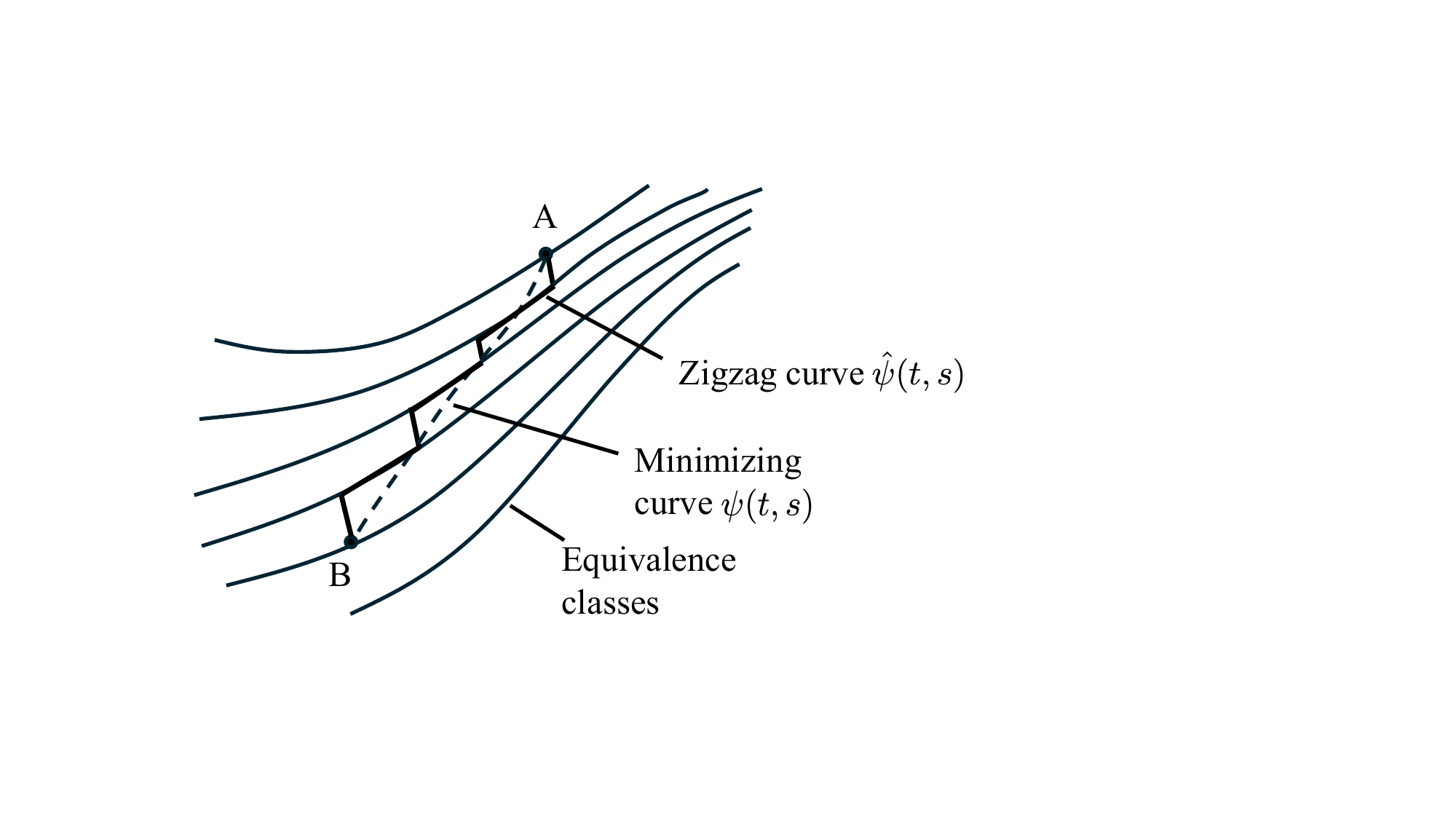}}
\caption{Illustration of a key step in proving the contraction of the quotient distance: a zigzag approximation of the minimizing curve and to prove contraction of every transverse zigs.}
\label{fig_quotient_distance}
\hfill\end{figure}

\noindent \textbf{2.} The LHS of (\ref{E:s1-2}) can be manipulated as
\begin{align}
    &\textstyle \frac{1}{2} \frac{d}{d t} \big\langle \mathcal P(t, \psi(t, s)) \frac{\partial\psi}{\partial s}(t, s), \mathcal P(t, \psi(t, s)) \frac{\partial\psi}{\partial s}(t, s) \big\rangle \notag \\
    &= \textstyle \big\langle \mathcal P(t, \psi(t, s)) \frac{\partial\psi}{\partial s}(t, s), \frac{d}{d t} \Big[\mathcal P(t, \psi(t, s)) \frac{\partial\psi}{\partial s}(t, s) \Big] \big\rangle \notag \\
    &= \textstyle \big\langle \mathcal P(t, \psi(t, s)) \frac{\partial\psi}{\partial s}(t, s), \mathcal P(t, \psi(t, s)) \frac{\partial}{\partial t} \frac{\partial\psi}{\partial s}(t, s) \big\rangle \notag \\
    &\quad\, \textstyle + \big\langle \mathcal P(t, \psi(t, s)) \frac{\partial\psi}{\partial s}(t, s), \frac{d}{d t} \mathcal P(t, \psi(t, s)) \frac{\partial\psi}{\partial s}(t, s) \big\rangle \notag \\
    &= \textstyle \big\langle \mathcal P(t, \psi(t, s)) \frac{\partial\psi}{\partial s}(t, s), \frac{\partial}{\partial \vect x} \Big[\mathcal P(t, \psi(t, s)) \frac{\partial}{\partial t} \psi(t, s) \Big] \frac{\partial\psi}{\partial s}(t, s) \big\rangle \notag \\
    &\quad\, \textstyle - \big\langle \mathcal P(t, \psi(t, s)) \frac{\partial\psi}{\partial s}(t, s), \frac{\partial}{\partial \vect x} \mathcal P(t, \psi(t, s)) \frac{\partial \psi}{\partial t}(t, s) \frac{\partial\psi}{\partial s}(t, s) \big\rangle \notag \\
    &\quad\, \textstyle + \big\langle \mathcal P(t, \psi(t, s)) \frac{\partial\psi}{\partial s}(t, s), \frac{d}{d t} \mathcal P(t, \psi(t, s)) \frac{\partial\psi}{\partial s}(t, s) \big\rangle \label{E:s1-9} \\
    &= \textstyle \big\langle \mathcal P(t, \psi(t, s)) \frac{\partial\psi}{\partial s}(t, s), \frac{\partial}{\partial \vect x} \Big[\mathcal P(t, \psi(t, s)) \frac{\partial}{\partial t} \psi(t, s) \Big] \frac{\partial\psi}{\partial s}(t, s) \big\rangle \notag \\
    &\quad\, \textstyle + \big\langle \mathcal P(t, \psi(t, s)) \frac{\partial\psi}{\partial s}(t, s), \frac{\partial}{\partial t} \mathcal P(t, \psi(t, s)) \frac{\partial\psi}{\partial s}(t, s) \big\rangle \label{E:s1-10} \\
    &= \textstyle \big\langle \mathcal P(t, \psi(t, s)) \frac{\partial\psi}{\partial s}(t, s), \frac{\partial}{\partial \vect x} \big[{-} \matr R \nabla H(t, \psi(t, s))\big] \frac{\partial\psi}{\partial s}(t, s) \big\rangle \notag \\
    &\quad\, \textstyle + \big\langle \mathcal P(t, \psi(t, s)) \frac{\partial\psi}{\partial s}(t, s), \frac{\partial}{\partial t} \mathcal P(t, \psi(t, s)) \frac{\partial\psi}{\partial s}(t, s) \big\rangle \label{E:s1-3} \\
    &\leq \textstyle - \Re\left\{ \Big[\mathcal P(t, \psi(t, s)) \frac{\partial\psi}{\partial s}(t, s)\Big]^\herm D^2 H(t, \psi(t, s)) \frac{\partial\psi}{\partial s}(t, s) \right\} \label{E:s1-4} \\
    &\leq \textstyle -c \Re\left\{ \Big[\mathcal P(t, \psi(t, s)) \frac{\partial\psi}{\partial s}(t, s)\Big]^\herm \matr R^{-1} \mathcal P(t, \psi(t, s)) \frac{\partial\psi}{\partial s}(t, s) \right\}. \label{E:s1-5} 
\end{align}
To obtain (\ref{E:s1-9}), we replaced $\frac{\partial}{\partial t} \frac{\partial \psi}{\partial s}(t, s)$ in the first term of the LHS by 
\begin{align*}
     \cdot = \frac{\partial}{\partial s} \frac{\partial \psi}{\partial t}(t, s) &= \frac{\partial}{\partial s} \vect f(t, \psi(t, s)) \\
     &= \frac{\partial}{\partial \vect x} \vect f(t, \psi(t, s)) \frac{\partial \psi}{\partial s}(t, s) \\
     &= \frac{\partial}{\partial \vect x} \frac{\partial}{\partial t} \psi(t, s) \frac{\partial \psi}{\partial s}(t, s).
\end{align*}
To obtain (\ref{E:s1-10}), we combined the last two terms in the LHS.
The reasoning for the (in-) equalities (\ref{E:s1-3}), (\ref{E:s1-4}), and (\ref{E:s1-5}) are as follows.
To obtain (\ref{E:s1-3}), the first term in (\ref{E:s1-10}) is simplified by substituting in
\begin{align*}
    \mathcal P(t, \vect x) \dot{\vect x} &= \mathcal P(t, \vect x) \big[\matr J(t, \vect x) \nabla H(t, \vect x) - \matr R \nabla H(t, \vect x) \big] \\
    &= -\matr R \nabla H(t, \vect x),
\end{align*}
where the last equality is a consequence of the definition (\ref{E:projection}).
%\begin{align*}
%    \dot{\vect x} - \mathcal P(\vect x) \dot{\vect x} &= \frac{\langle \matr J(\vect x) \nabla H(\vect x), \matr R^{-1} \dot{\vect x} \rangle}{\|\dot{\vect x} \| \| \matr J(\vect x) \nabla H(\vect x) \|} \matr J(\vect x) \nabla H(\vect x) \\
%    &= \frac{\langle \matr J(\vect x) \nabla H(\vect x), -\nabla H(\vect x) \rangle}{\|\dot{\vect x} \| \| \matr J(\vect x) \nabla H(\vect x) \|} \matr J(\vect x) \nabla H(\vect x) = 0.
%\end{align*}
In the LHS of (\ref{E:s1-4}), the second term is eliminated as follows. Note that, the integral of the second term over $s\in [0, 1]$, can be approximated up to arbitrary accuracy with a series of zigzag curve segments $\hat\psi_i(t, s),\, i = 1,\ldots,\, N$ such that
\begin{equation*}
    \mathcal P(t, \hat\psi_i(t, s)) \frac{\partial\hat\psi_i}{\partial s}(t, s) = \begin{cases}
        \frac{\partial\hat\psi_i}{\partial s}(t, s) &\text{transverse zig} \\
        0 &\text{parallel zag}
    \end{cases}
\end{equation*}
The approximation is illustrated in Fig~\ref{fig_quotient_distance}.
Since $\mathcal P(t, \hat\psi_i(t, s))$ shrinks the transverse zigs for $t > t_0$, for the transverse zigs, we have that
\begin{align*}
    &\Big\langle \mathcal P(t, \hat\psi_i(t, s)) \frac{\partial\hat\psi_i}{\partial s}(t, s), \frac{\partial}{\partial t} \mathcal P(t, \hat\psi_i(t, s)) \frac{\partial\hat\psi_i}{\partial s}(t, s) \Big\rangle \\
    &= \Big\langle \frac{\partial\hat\psi_i}{\partial s}(t, s), \frac{\partial}{\partial t} \mathcal P(t, \hat\psi_i(t, s)) \frac{\partial\hat\psi_i}{\partial s}(t, s) \Big\rangle\leq 0,
\end{align*}
because any change in the projection $\mathcal P(t, \hat\psi_i(t, s))$ decreases the length of the zig segment,
and, for the parallel zags,
\begin{align*}
    &\Big\langle \mathcal P(t, \hat\psi_i(t, s)) \frac{\partial\hat\psi_i}{\partial s}(t, s), \frac{\partial}{\partial t} \mathcal P(t, \hat\psi_i(t, s)) \frac{\partial\hat\psi_i}{\partial s}(t, s) \Big\rangle \\
    &= \Big\langle 0_n, \frac{\partial}{\partial t} \mathcal P(t, \hat\psi_i(t, s)) \frac{\partial\hat\psi_i}{\partial s}(t, s) \Big\rangle = 0
\end{align*}
because $\mathcal P(t, \hat\psi_i(t, s)) \frac{\partial\hat\psi_i}{\partial s}(t, s) = 0_n$.
To obtain (\ref{E:s1-5}), we used the same zigzag approximation and
\begin{equation*}
    c\, \lambda_{\max}(\matr R^{-1}) = \frac{c}{\lambda_{\min}(\matr R)} = \lambda_{\min}(D^2 H(t, \vect x)).
\end{equation*}

\noindent \textbf{3.} From steps 1 and 2, we have proved that, at $t = t_0$ and for a series of zigzag curve segments $\hat\psi_i(t, s),\, 1 = 1,\ldots,\, N$ approximating the minimizing (at $t = t_0$) curve $\psi(t, s)$, the inequality (\ref{E:s1-6}) holds. Since as the number of zigzag curves segments $N$ increases, both the LHS and the RHS of (\ref{E:s1-6}) converge to the minimum value. Hence (\ref{E:s1-1}) holds. \hfill $\square$

\subsection{Proof of Proposition~\ref{prop_no_diss}}
Choose the inner product $\langle \vect y, \vect x \rangle = \Re\{\vect y^* \vect x\}$. The proof is otherwise the same as the proof of Proposition~\ref{prop_main}. \hfill $\square$

\subsection{Proof of Proposition~\ref{prop_limit_cycle}}
From Proposition~\ref{prop_main}, we have that, for every initial condition $(t_0, \vect x_0)$, there is
\begin{equation*}
    \lim_{t\to \infty} \dist(t, \Phi(t, t_0, \vect x_0), \eqm{\vect x}(t)) = 0.
\end{equation*}
By property (iv) of the quotient distance that follows (\ref{E:quotient_distance}), it implies that
\begin{equation*}
    \lim_{t\to \infty} H(t, \Phi(t, t_0, \vect x_0) - H(t, \eqm{\vect x}(t)) = 0.
\end{equation*}
This completes the proof. \hfill $\square$

\subsection{Proof of Proposition~\ref{prop_convergence}}
From Proposition~\ref{prop_limit_cycle}, it suffices to show that there is a particular solution $\eqm{\vect x}(t)$ such that
\begin{equation} \label{E:s6-1}
    \frac{d}{dt} H(t, \eqm{\vect x}(t)) = 0
\end{equation}
for all $t \in \mathbb{R}$. To this end, consider the set $E = E_t$ in which the Hamiltonian has zero derivative. By definition, $E$ is an invariant set. Assume that the flow is complete. Then from any initial condition $(t_0, \vect x_0) \in \mathbb{R} \times E$, the solution $\eqm{\vect x}(t) = \Phi(t, t_0, \vect x_0)$ satisfies (\ref{E:s6-1}). Hence we have found a particular solution, which completes the proof. \hfill $\square$

\subsection{Proof of Lemma~\ref{lem_RLC}}

We change to coordinates that are rotating at the frequency $\omega_0$ by the change of variables
\begin{equation*}
    \vect x \leftarrow e^{-j\omega_0 t} \vect x.
\end{equation*}
The system equation then writes
\begin{equation*}
    \dot{\vect x} = (\matr J_1 - \matr R) \nabla H(\vect x) + \matr G u_1
\end{equation*}
where $u_1 = 1$ and $\matr J_1 = \matr J - j \omega_0 \matr Q^{-1}$. 
Now, consider the shifted Hamiltonian function
\begin{equation*}
    \mathcal H(\vect x, \eqm{\vect x}(0)) = \frac{1}{2} \big[\vect x - \eqm{\vect x}(0)\big]^\herm \matr Q \big[\vect x - \eqm{\vect x}(0)\big].
\end{equation*}
It is easy to find its time derivative as~\cite{monshizadeh2019conditions}
\begin{align*}
    \dot{\mathcal H}(\vect x, \eqm{\vect x}(0)) &= -\big[\vect x - \eqm{\vect x}(0)\big]^\herm \matr Q \matr R \matr Q \big[\vect x - \eqm{\vect x}(0) \big] \\
    &\leq -\lambda_{\min}(\matr R)\, \lambda_{\min}(\matr Q)\, \mathcal H( \vect x, \eqm{\vect x}(0)).
\end{align*}
Hence
\begin{align*}
    &\lim_{t\to\infty} \mathcal H(\vect x(t), \eqm{\vect x}(0)) \\
    &= \lim_{t\to \infty} \frac{1}{2} \big[ \vect x(t) - \eqm{\vect x}(0)\big]^\herm \matr Q \big[ \vect x(t) - \eqm{\vect x}(0)\big] = 0.
\end{align*}
Hence we obtain that the state vector $\vect x$ converges to the limit cycle $\eqm{\vect x}(0)$, and the orbit of the limit cycle is the limit set of all solutions. \hfill $\square$
%Note that, as a more granular distributed-parameter model is used, i.e., as $\varepsilon \to 0$, the co-state $\varepsilon \vect x = \nabla H(\vect x)$ remains to have the same order of magnitude as the voltages and currents. Hence we obtain that the co-state $\varepsilon \vect x$ converges to the limit cycle uniformly in the granular level of distributed-parameter model $\varepsilon$, and the orbit of the limit cycle is the limit set of all solutions.

\subsection{Proof of Proposition~\ref{prop_final}}

%Since the Hamiltonian, $H(\vect s) = \frac{1}{2} \vect s^\herm \matr Q^{-1} \vect s$, expressed in the co-state coordinates is a quadratic form, the limit cycle $\eqm{\vect x}(\tau)$ is contained in a level set of $H(\vect s)$.
By Proposition~\ref{prop_convergence}, we obtain that the limit set of every solution is contained in the level set of $H(\vect s) = \frac{1}{2} \vect s^\herm \matr Q^{-1} \vect s$ occupied by $\eqm{\vect x}(t)$, which is the first constraint we will use to characterize the limit set. We proceed to prove that the Hamiltonian of every edge satisfies the same property.

Consider a perturbation of the gradient of the Hamiltonian written as
\begin{equation*}
    \nabla \hat H(t, \vect x_1) = \matr P^{-1} \nabla H(t, \vect x_1)
\end{equation*}
where
\begin{equation*}
    \matr P = \mathrm{diag}(p_1 \matr I_2,\, p_2 \matr I_2,\, p_3,\, p_4,\, p_5,\, p_6,\, p_7 )
\end{equation*}
for $p_i > 0,\, i = 1,\ldots, 7$. The overall pH system can then be written as
\begin{equation*}
    \dot{\vect x}_1 = (\matr J(\vect x) - \matr R ) \matr P \nabla \hat H(t, \vect x_1).
\end{equation*}
It can be checked that
\begin{equation} \label{E:s3-1}
    \frac{1}{2} \left[(\matr J(\vect x) + \matr R ) \matr P + \matr P (\matr J(\vect x) + \matr R)^\herm\right]
\end{equation}
is a constant matrix, and (\ref{E:s3-1}) remains negative definite if we choose $\col(p_i) \approx 1_7$. By Proposition~\ref{prop_limit_cycle}, we have that the value of $\hat H(\vect s)$ on the positive limit set should be equal to the value at $\eqm{\vect x}(\tau)$. By choosing linearly independent $\col(p_i)$'s we can fix the value of the Hamiltonian of every edge; that is, the voltage amplitude of every shunt capacitor, the current amplitude of every R--L line, and the energy stored in every SG are all equal to their values at $\eqm{\vect x}(\tau)$.

To separate the Hamiltonian associated with the mechanical and the electrical energy of the SG, consider a perturbed system having a shunt capacitor that splits the stator inductance of each SG; that is, the subsystem (the subscript $i \in \{ 1, 2\}$ is omitted)
\begin{equation*}
    L \dot I = -R I - \psi j \omega e^{j\theta} + V
\end{equation*}
is replaced by
\begin{equation*}
    \begin{cases}
        \alpha L \dot I_1 = -\alpha R I_1 - \psi j \omega e^{j\theta} + V_1 \\
        C \dot V_1 = -G V_1 + I_2 - I_1 \\
        (1 - \alpha) L \dot I_2 = -(1 - \alpha) R I_2 - V_1 + V
    \end{cases}
\end{equation*}
where $0 < \alpha < 1$ and $C, G > 0$ are small.
By Theorem~3.5 in~\cite{Khalil:1173048}, as $\alpha, C, G$ tend to zero, the solutions of the perturbed system tends to those of the original system. 
Applying the same contraction analysis to the perturbed system, we have that the mechanical energy $\frac{1}{2} J \omega_i^2$ and stator electrical energy $\frac{1}{2} L_i \|I_i\|^2$ are both equal to their value at the perturbed limit cycle. Taking the added shunt capacitance to zero, we obtain that the value of $\omega_i$ in the positive limit set of the original system is equal to the synchronized frequency of $\eqm{\vect x}(\tau)$. The dynamics on the positive limit set is then constrained to be a passive RLC circuit with two voltage sources of the same frequency. By Lemma~\ref{lem_RLC}, the orbit of $\eqm{\vect x}(\tau)$ is the only possible limit set. \hfill $\square$

\bibliographystyle{IEEEtran}
\bibliography{References}

\end{document}